\documentclass{article}

\usepackage[english]{babel}

\usepackage[letterpaper,top=2cm,bottom=2cm,left=3cm,right=3cm,marginparwidth=1.75cm]{geometry}

\usepackage{amsmath}
\usepackage{graphicx}
\usepackage{amssymb}
\usepackage[colorlinks=true, allcolors=blue]{hyperref}
\usepackage{booktabs}
\usepackage[flushleft]{threeparttable}
\usepackage{csquotes}
\usepackage[backend=biber,style=apa]{biblatex}
\title{Cross-Sectional Heterogeneity in LSTM Networks for Financial Time Series}
\author{Julius Döbelt \\[2pt]
\normalsize Technical University of Darmstadt \\
\normalsize Department of Law and Economics \\
\normalsize Hochschulstraße 1 \\
\normalsize D-64289 Darmstadt, Germany \\
\texttt{\normalsize julius.doebelt@tu-darmstadt.de}}

\date{August 2026}

\begin{document}
\maketitle
\begin{abstract}
Predicting financial asset returns remains one of the most difficult challenges in empirical finance, driven by the low signal-to-noise ratio and the semi-strong form of market efficiency. While deep learning models, especially Long Short-Term Memory (LSTM) networks, have shown promise in capturing temporal dependencies, standard architectures often struggle to account for the cross-sectional heterogeneity of asset returns. This paper proposes a novel architectural extension to the basic LSTM model designed to improve both predictive accuracy and model interpretability. The framework integrates macro-financial covariates to capture broader economic signals and learnable sector embeddings to encompass heterogeneity by sector. Furthermore, regularization techniques like label smoothing, dropout and gradient clipping are used to mitigate the risk of overfitting to the training distribution. The trading strategy involves constructing a long-short portfolio based on daily directional forecasts for each S\&P 500 constituent, targeting stocks expected to under- or outperform the cross-sectional median return of the S\&P 500. Model Performance is evaluated against three competitive benchmarks: a basic LSTM, a Random Forest model and a traditional market buy-and-hold strategy. The empirical results demonstrate that the LSTM with sector embeddings outperforms all benchmarks across key risk and return metrics. By utilizing sector embeddings, the model explicitly incorporates cross-sectional heterogeneity, allowing it to adapt to varying industry dynamics within the market. To address the black-box nature of deep learning, I use latent space visualizations to analyse how the model differentiates between sectors, providing insights into the internal representation of the sectors in the LSTM. The impact of the sector information can be quantified using a novel contribution metric by inspecting the weights of the LSTM. The predictive signal is driven by a short-term reversal factor and an industry momentum factor. (JEL C45, C53, C58, G11)
\end{abstract}


\section{Introduction}

The prediction of financial asset returns occupies a central position in empirical finance because it is simultaneously among the most extensively studied and yet unresolved problems in the literature. The efficient market hypothesis posits that publicly available information is rapidly incorporated into asset prices, leaving little exploitable predictive content in historical data \parencite{famaEfficientCapitalMarkets1970, famaEfficientCapitalMarkets1991}. Yet decades of empirical work have documented systematic deviations from this hypothesis, organized either as cross-sectional anomalies tied to firm characteristics \parencite{famaCommonRiskFactors1993, carhartPersistenceMutualFund1997, famaFivefactorAssetPricing2015} or as weak but persistent time-series predictability of the equity premium \parencite{campbellPredictingExcessStock2008}. From the literature emerged consensus that return predictability is regime-dependent, and easily overstated through specification searches and p-hacking \parencite{welchComprehensiveLookEmpirical2008, harveyCrossSectionExpectedReturns2016}. Two features make the problem particularly demanding: an exceptionally low signal-to-noise ratio in daily returns and a predictor space whose dimensionality grows rapidly relative to the effective sample size, raising concerns about overfitting and false discovery \parencite{israelCanMachinesLearn2020}. These characteristics make return prediction a challenging testbed for modern machine learning. Traditional econometric methods, while transparent and theoretically grounded, struggle with nonlinear interactions, high-dimensional predictor spaces, and time-varying relationships \parencite{guEmpiricalAssetPricing2020}. Linear factor models impose strong functional form assumptions that may obscure economically meaningful interactions. The rise of hundreds of potential predictors, the so-called factor zoo, has made specification of linear models for equity risk premium prediction less straightforward \parencite{fengTamingFactorZoo2020, harveyCrossSectionExpectedReturns2016}. Machine learning methods offer a natural complement: they can accommodate large predictor sets, learn nonlinear mappings, and exploit regularization to manage the bias–variance trade-off in noisy environments \parencite{guEmpiricalAssetPricing2020}. The empirical evidence assembled over the past five years suggests that these tools deliver meaningful gains in out-of-sample $R^{2}$ and economic value, though the magnitude of the gains and the difficulty to exploit the strategies remain subjects of active debate \parencite{avramovMachineLearningVs2021}.

\subsection{Deep Learning in Finance}

Within the broader field of machine learning, neural networks have attracted particular attention for their flexibility in representing complex functional forms without explicit feature engineering \parencite{goodfellow2016}. Their adoption in asset pricing has accelerated since the seminal contribution \textcite{guEmpiricalAssetPricing2020} who systematically benchmark a range of machine learning methods, including penalized linear models, regression trees, and feed-forward neural networks, for monthly U.S. equity return prediction. They document that neural networks deliver the strongest performance, more than doubling the out-of-sample $R^{2}$ of leading regression methods, and that the economic value of these gains is concentrated in difficult-to-arbitrage securities. \textcite{chenDeepLearningAsset2024} show that by embedding deep learning directly within a no-arbitrage stochastic discount factor framework, showing that recurrent architectures combined with generative adversarial training can capture rich conditional information that static models miss. \textcite{bianchiBondRiskPremiums2021} document analogous gains for the U.S. bond market, where neural networks dominate linear benchmarks in forecasting bond risk premia. A parallel strand of literature has focused specifically on temporal architectures designed to exploit sequential dependencies in financial time series. Recurrent neural networks (RNNs) extend the feed-forward paradigm by introducing internal state that propagates across time steps, allowing the network to condition predictions on the full history of an input sequence. In practice, vanilla RNNs are difficult to train over long sequences due to the vanishing and exploding gradient problem. The Long Short-Term Memory (LSTM) architecture of \textcite{hochreiterLongShortTermMemory1997} addresses this limitation through a gated memory cell that permits gradient flow over long horizons, with subsequent refinements by \textcite{gersLearningForgetContinual2000} and a systematic empirical evaluation of LSTM variants by \textcite{greffLSTMSearchSpace2017}. Gated recurrent units \parencite{choPropertiesNeuralMachine2014} provide a streamlined alternative with comparable empirical performance. The application of LSTMs to financial prediction tasks was popularized by \cite{fischerDeepLearningLong2018}, who deploy a single-layer LSTM on the constituents of the S\&P 500 and report substantial annualized returns prior to 2010, with a marked deterioration thereafter. Their methodology builds on the earlier benchmarking exercise of \textcite{kraussDeepNeuralNetworks2017}, who compare deep neural networks, gradient-boosted trees, and random forests for statistical arbitrage on the same universe. They find broadly comparable performance between the models while noting that hyperparameter tuning is essential for neural networks. Beyond directional prediction, \textcite{sirignanoUniversalFeaturesPrice2019} train deep recurrent networks on a large sample of limit order book data and document a universal component of price formation that transfers across stocks, a finding that anticipates the cross-sectional structure exploited in the present paper. \textcite{baoDeepLearningFramework2017} integrate wavelet denoising with stacked LSTMs for index-level forecasting, while \textcite{krausDecisionSupportFinancial2017} combine LSTM-based text representations with returns to predict price movements following corporate disclosures. Comprehensive surveys are provided by \textcite{sezerFinancialTimeSeries2020} and, with an asset-pricing emphasis, by \cite{israelCanMachinesLearn2020}. Despite this rapid expansion of the literature, several methodological tensions remain unresolved. First, the predominant practice of training a single sequence model on pooled cross-sectional data, as in \cite{fischerDeepLearningLong2018}, implicitly assumes that all assets are drawn from a common return-generating process. This assumption is at odds with a large body of evidence that industry membership, size, and other firm characteristics induce systematic cross-sectional heterogeneity in return dynamics \parencite{moskowitzIndustriesExplainMomentum1999, hongIndustriesLeadStock2007}. Second, the deep learning literature in finance has been slow to absorb the architectural innovations developed in natural language processing, where learnable embeddings of categorical entities have become standard for incorporating discrete features \parencite{bengioNeuralProbabilisticLanguage2003, mikolovDistributedRepresentationsWords2013}. Third, the strong returns reported in early LSTM studies \parencite{fischerDeepLearningLong2018} have not been systematically validated over more recent market regimes, leaving open the question of whether the documented decline after 2010 reflects a durable shift in market efficiency or a temporary phenomenon.

\subsection{Industry Momentum}

The relevance of industry-level structure for return prediction has a long and well-established history in asset pricing. \cite{jegadeeshReturnsBuyingWinners1993} document that strategies buying past winners and selling past losers over horizons of three to twelve months generate economically and statistically significant abnormal returns, a finding subsequently confirmed by \textcite{jegadeeshProfitabilityMomentumStrategies2001} and extended internationally by \cite{asnessSizeMattersIf2018}. \cite{moskowitzIndustriesExplainMomentum1999} provide the foundational result for the present paper: a substantial share of individual stock momentum profits is attributable to industry momentum, with high-momentum industries outperforming low-momentum industries by economically meaningful amounts over the subsequent six months. \textcite{grundyUnderstandingNatureRisks2001} further decompose the momentum return and show that industry-level autocorrelation is a primary source of the strategy's profitability, in contrast to the short-horizon return reversal that characterizes individual-stock momentum \parencite{lewellenMomentumAutocorrelationStock2002}. The robustness of industry momentum is, however, conditional on market regime. \cite{danielMomentumCrashes2016} document that momentum strategies experience occasional but severe crashes, typically following sharp market rebounds in which prior losers, overrepresented in the short leg of the strategy, rebound disproportionately. Fourteen of the fifteen worst monthly momentum returns in their sample occur when the two-year lagged market return is negative, and the resulting drawdowns can exceed those of the underlying market. The crashes of 2001–2002 and 2009 are emblematic in this regard. Related work by \textcite{hongIndustriesLeadStock2007} establishes that information diffuses across industries with measurable lags, providing a structural mechanism through which industry-level signals can predict future returns even in informationally efficient markets. At shorter horizons, the literature has documented robust short-term reversal effects: stocks with strong returns over the prior week tend to underperform in the subsequent week, and vice versa \parencite{lehmannFadsMartingalesMarket1990, jegadeesh1990evidence}. The coexistence of medium-horizon industry momentum and short-horizon individual reversal therefore implies that a model operating on daily data must, implicitly or explicitly, distinguish between these two opposing temporal signals. These stylized facts motivate an architecture in which industry membership enters the model as an explicit, learnable representation rather than as an input feature that the network must infer from raw returns. \textcite{guoEntityEmbeddingsCategorical2016} demonstrate that this approach extends naturally to structured tabular data, where embeddings of high-cardinality categorical variables outperform one-hot encodings in both predictive accuracy and parameter efficiency.

\subsection{Explainability of Deep Learning Models}

A persistent concern in the application of deep learning to economic and financial problems is the limited interpretability of the resulting models \parencite{rudinStopExplainingBlack2019, murdochDefinitionsMethodsApplications2019}. Whereas the coefficients of a linear factor model admit a direct economic interpretation as risk loadings or anomaly exposures, the thousands or even millions of parameters in a deep neural network do not lend themselves to such transparent reading. The interpretability literature has responded with two broad approaches. Post-hoc methods, most prominently LIME \parencite{ribeiroWhyShouldTrust2016} and SHAP \parencite{lundbergUnifiedApproachInterpreting2017}, attempt to attribute predictions of a trained model to its inputs through local approximations or game-theoretic value functions. Intrinsically interpretable models, by contrast, are designed from the outset to admit transparent inspection of their learned representations \parencite{rudinStopExplainingBlack2019}. In the context of finance, where the cost of model failure is borne directly by investors and where regulators increasingly demand explanation \parencite{brackeMachineLearningExplainability2019}, the case for interpretability is particularly strong. Within deep learning, this has motivated architectures such as neural additive models \parencite{agarwal2021neural}, which learn a separate subnetwork per feature so that each feature's contribution can be inspected directly; concept bottleneck models \parencite{koh2020concept}, which route predictions through a layer of human-specified concepts; and prototype-based networks \parencite{chen2019looks}, which classify by comparison to learned exemplars.

\bigskip

This paper makes three primary contributions to the literature. First, I extend the single-layer return-only LSTM of \textcite{fischerDeepLearningLong2018} along three architectural dimensions: the integration of macro-financial covariates, the extensive usage of regularization, and most centrally the inclusion of learnable sector embeddings based on TRBC (The Refinitiv Business Classification). The embedding architecture is the first application, to my knowledge, of a continuous representation of industry membership within a recurrent network for cross-sectional return prediction. It introduces cross-sectional heterogeneity into a model class that has historically treated assets as exchangeable. Second, I provide a longer-term validation of LSTM performance in modern market regimes. \textcite{fischerDeepLearningLong2018} document a marked deterioration of LSTM-based statistical arbitrage profits after 2010, but their sample ends in 2015. By extending the evaluation through the end of 2024, covering an additional decade that includes the COVID-19 shock, the post-pandemic rate normalization, and the recent inflation regime, I assess whether the post-2010 decline reflects a durable shift in market efficiency or a transient phenomenon. The results confirm the deterioration documented by \textcite{fischerDeepLearningLong2018} and extend it: with the partial exception of 2020, no year after 2008 produces strategy returns approaching those of the late 1990s, consistent with the broader literature on diminishing returns to quantitative trading strategies. Third, I open the black box to a meaningful extent by decomposing the predictive signal of the sector LSTM into its temporal and cross-sectional components. I show that the learned sector embeddings function as an implicit industry momentum factor. The temporal component of the model, in turn, displays the short-horizon reversal pattern documented in \textcite{lehmannFadsMartingalesMarket1990} and \textcite{jegadeesh1990evidence}. 

The remainder of the paper is organized as follows. Section 2 develops the methodology, beginning with the rolling-window evaluation protocol and proceeding to the LSTM architecture and the three proposed extensions. Section 3 describes the data, including the construction of a survivorship-bias-free S\&P 500 universe and the macro-financial covariate set. Section 4 presents the empirical results: performance against benchmarks, the long-term validation, and the decomposition of the predictive signal into known factors. Furthermore, the robustness of the LSTM models are analyzed. Section 6 concludes and outlines directions for further work.

\section{Methodology}

To ensure the comparability of the results with the baseline established by \textcite{fischerDeepLearningLong2018}, I adopt their rolling-window approach for data segmentation and feature generation. However, I deviate slightly in certain data-preprocessing steps to account for recent advances in the literature on Neural Networks which are described in the subsequent parts of this section.

\subsection{Training and Test Set}

Following the standard procedure for walking-forward validation in financial time series, the dataset is divided into 27 study periods. Each study period constitutes a rolling block of four years, split into a training set of three years and a test set of one year. One year consists of roughly 252 trading days on average.

The training process utilizes sliding windows within the first three years. Specifically, I use a look-back sequence of 60 days to predict the directional movement of the subsequent day. Once the model parameters are optimized on the training set, the model is deployed on the subsequent one year test set in a strictly out-of-sample fashion. The two main modification to the methodology from \textcite{fischerDeepLearningLong2018} are therefore: 1) Using only 60 instead of 240 lags of stock returns as input to the LSTM, that avoids overlap in the sequences from the train and test set. \textcite{fischerDeepLearningLong2018} used 750 days for training, where 20\% (or 150 days) are used for the validation set. This means that there is still an overlap of 90 past stock returns for the last date in the training data and first date in the test data. This approach has the issue of data leakage as the training set contains information that is also in the testing set \parencite{prado10ReasonsMost2018}. Additionally, I use only 60 lag-sequences as input, because more lags make it harder for the LSTM to converge \parencite{huOvercomingVanishingGradient2019}. 2) Using exactly three years of training data and one year of test data. This is more sensible when plotting results grouped by year, as these predictions then come from one model only, whose properties can be inspected as I provide the model weights for each of the models used in this study. This facilitates the analysis for certain years when the predictions are especially strong or weak. 

\begin{figure}
\centering
\includegraphics[width=1\linewidth]{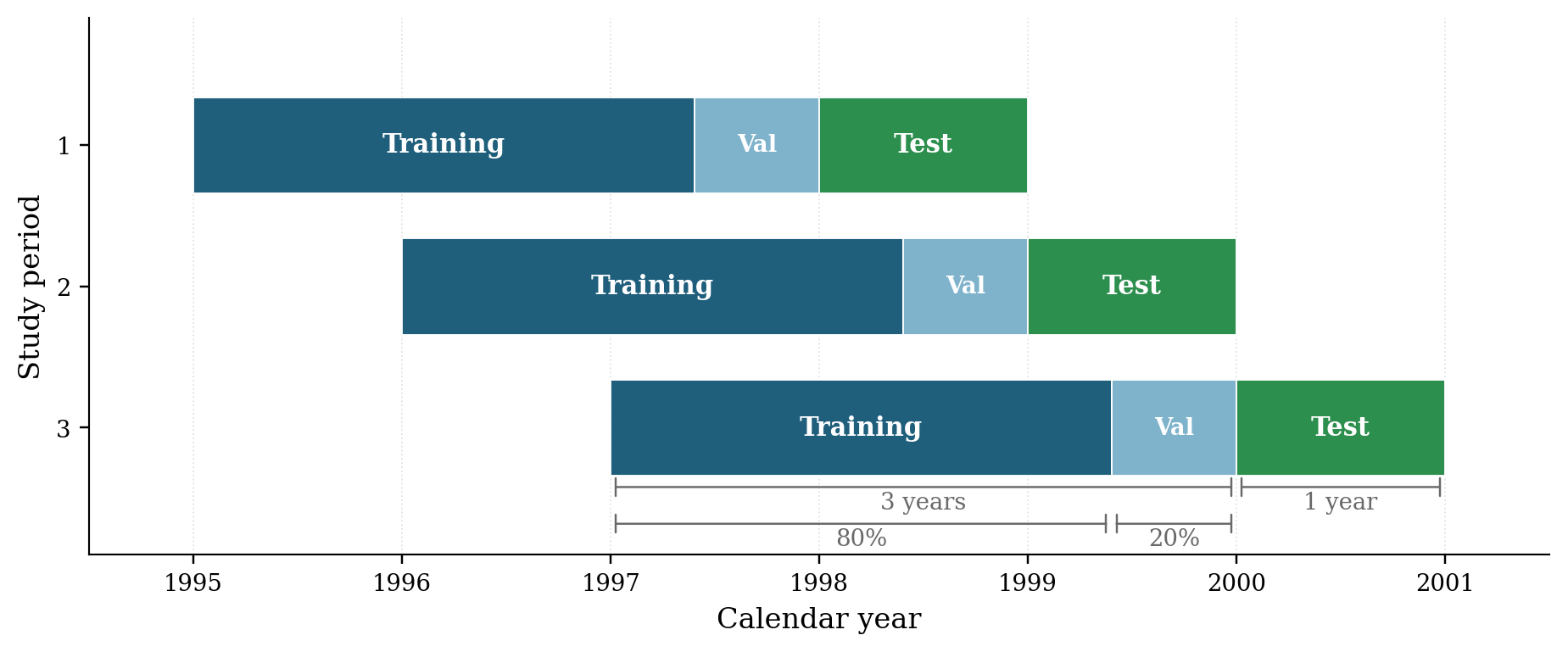}
\caption{\label{fig:traintest}Data Split into Training, Validation and Test for First 3 Study Periods.}
\end{figure}

After the completion of one study period, the entire four-year block is rolled forward by one year. This process yields 27 non-overlapping trading periods, which allows to evaluate performance across different market regimes without look-ahead bias. Figure \ref{fig:traintest} illustrates the three periods training, validation and test for the first study periods.

I consider all stocks that are constituents of the S\&P 500 at the end of the training period. To eliminate survivor bias, I include all stocks available in the training history and retain them in the trading set as long as price data is available, even if they are delisted during the trading period.

\subsection{Forecasting Task}

Let $P_{t}^{s}$ denote the price of stock s at time t, where $s\in\{1,...,n_{t}\}$ and $n_{t}$ is the number of stocks in the S\&P 500 at time t. I calculate the simple one-day return $R_{t}^{s}$ as:

$$R_{t}^{s}=\frac{P_{t}^{s}-P_{t-1}^{s}}{P_{t-1}^{s}}$$

To facilitate the training of the neural network and ensure convergence, the returns are standardized. Crucially, the mean ($\mu_{train}$) and standard deviation ($\sigma_{train}$) used for standardization are derived solely from the training set to prevent any data leakage or look-ahead bias. The standardized return $\tilde{R}_{t}^{s}$ is given by:

$$\tilde{R}_{t}^{s}=\frac{R_{t}^{s}-\mu_{train}}{\sigma_{train}}$$

The standardized returns are stacked into one large vector of size $n_{i}\cdot T_{study}$, where $T_{study}$ are the number of days in the study period, ordered by date and $n_i$ is the number of stocks that are considered for study period $i$. The LSTM expects input of sequences. Hence, the primary input features are sequences of standardized daily returns. In the base model, the only input to the LSTM is the sequence of 60 lagged stock returns. For the prediction of $\tilde{R}_{t+1}^{s}$, the past 60 returns ${\tilde{R}_{t-59}^{s},\tilde{R}_{t-58}^{s},...,\tilde{R}_{t}^{s}}$ are used.

The total number of observation are approximately 465,000 varying slightly with the number of stocks in S\&P 500, of which about 340,000 are used for training and validation and about 125,000 are used for the test data.

The forecasting task is framed as a binary classification problem. For each trading day, the median stock return is calculated from the cross-sectional stock returns of all the stocks in the S\&P 500. Stock returns are put into two classes, above median or below median stock returns. The goal is to predict the class for each stock on every trading day. This approach transforms the regression problem of absolute price movements into a directional signal more suitable for a long-short investment strategy \parencite{huckPairsSelectionOutranking2009}. The daily trading strategy consists of going long with the k stocks that have highest probability for class 1 (outperforming S\&P 500) and going short the k stocks with highest probability for class 0 (underperforming S\&P 500). This results in a long-short portfolio with $2k$ stocks. The general prediction setup is similar to \textcite{fischerDeepLearningLong2018}.

\subsection{LSTM}

\begin{figure}
\centering
\includegraphics[width=0.6\linewidth]{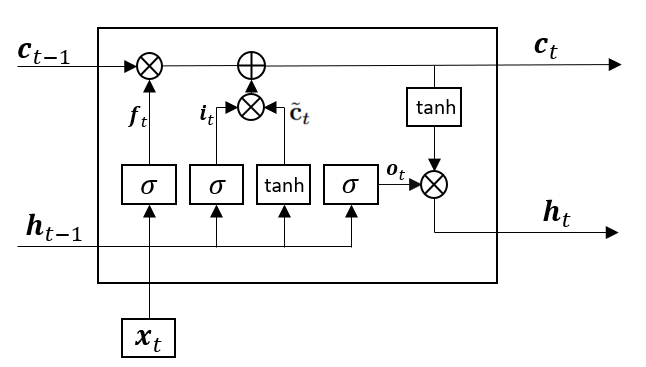}
\caption{\label{fig:lstmCell}LSTM Cell.}
\end{figure}

LSTM's are Recurrent Neural Networks that can overcome the vanishing and exploding gradient problem, that is a main limitation in plain RNN's. They were invented by \textcite{hochreiterLongShortTermMemory1997} to learn long-range dependencies in sequences which makes them a useful method in time series forecasting. There have been refinements including (but not limited to) by \textcite{gersLearningForgetContinual2000}, \textcite{gravesFramewisePhonemeClassification2005a} and \textcite{greffLSTMSearchSpace2017} and new methods developed based on the LSTM architecture, most popular the Gated Recurrent Unit (GRU) \parencite{choPropertiesNeuralMachine2014}. Effectiveness of LSTM's was shown in many fields, e.g. in Finance \parencite{fischerDeepLearningLong2018}, Computer Science \parencite{sutskeverSequenceSequenceLearning2014} and Hydrology \parencite{kratzertRainfallRunoffModelling2018}. For a light introduction, see \textcite{karpathyUnreasonableEffectivenessRecurrent2020} and \textcite{olahUnderstandingLSTMNetworks} which I also use as source for describing the LSTM subsequently. Figure \ref{fig:lstmCell} visualizes the architecture of the LSTM Cell.

The core innovation of the LSTM is the cell state $\mathbf{c}_{t}$, an internal memory that can carry information across long sequences with minimal attenuation. The interaction with this memory depends on three learnable gates. The Forget Gate $\mathbf{f}_{t}$ decides what information from the previous cell state $\mathbf{c}_{t-1}$ is no longer relevant and should be discarded. $$\mathbf{f}_{t}=\sigma(W_{f}\cdot[\mathbf{h}_{t-1},\mathbf{x}_{t}]+\mathbf{b}_{f})$$ $\sigma$  is the sigmoid activation function, mapping values to the interval (0,1). W and b are the learnable weight matrices and bias vectors, respectively. If $\mathbf{f}_{t}=0_H$, the old memory is completely forgotten, if $\mathbf{f}_{t}=1_H$, it is passed through entirely. The Input Gate $i_{t}$ and Candidate State $\tilde{\mathbf{c}}_{t}$ determine what new information will be stored in the cell state. The sigmoid gate $\sigma$ decides which values to update, while the $\tanh$  layer creates a vector of new candidate values. 
$$\mathbf{i}_{t}=\sigma(W_{i}\cdot[\mathbf{h}_{t-1},\mathbf{x}_{t}]+\mathbf{b}_{i})$$
$$\tilde{\mathbf{c}}_{t}=\tanh(W_{\tilde{c}}\cdot[\mathbf{h}_{t-1},\mathbf{x}_{t}]+\mathbf{b}_{\tilde{c}})$$
The new memory $c_{t}$ is a combination of the filtered old memory and the scaled new candidates. This linear update is the key to preventing vanishing gradients. $\odot$ denotes the Hadamard product.
$$\mathbf{c}_{t}=\mathbf{f}_{t}\odot \mathbf{c}_{t-1}+\mathbf{i}_{t}\odot\mathbf{\tilde{c}}_{t}$$ 
The main vector of interest for the prediction task is the hidden state vector $\mathbf{h}_{t}$ which is a filtered version of the updated memory $\mathbf{c}_{t}$ and protected by the output gate $\mathbf{o}_t$. $$\mathbf{o}_{t}=\sigma(W_{o}\cdot[\mathbf{h}_{t-1},\mathbf{x}_{t}]+\mathbf{b}_{o})$$
$$\mathbf{h}_{t}=\mathbf{o}_{t}\odot\tanh(\mathbf{c}_{t})$$ 

\begin{figure}
\centering
\includegraphics[width=1\linewidth]{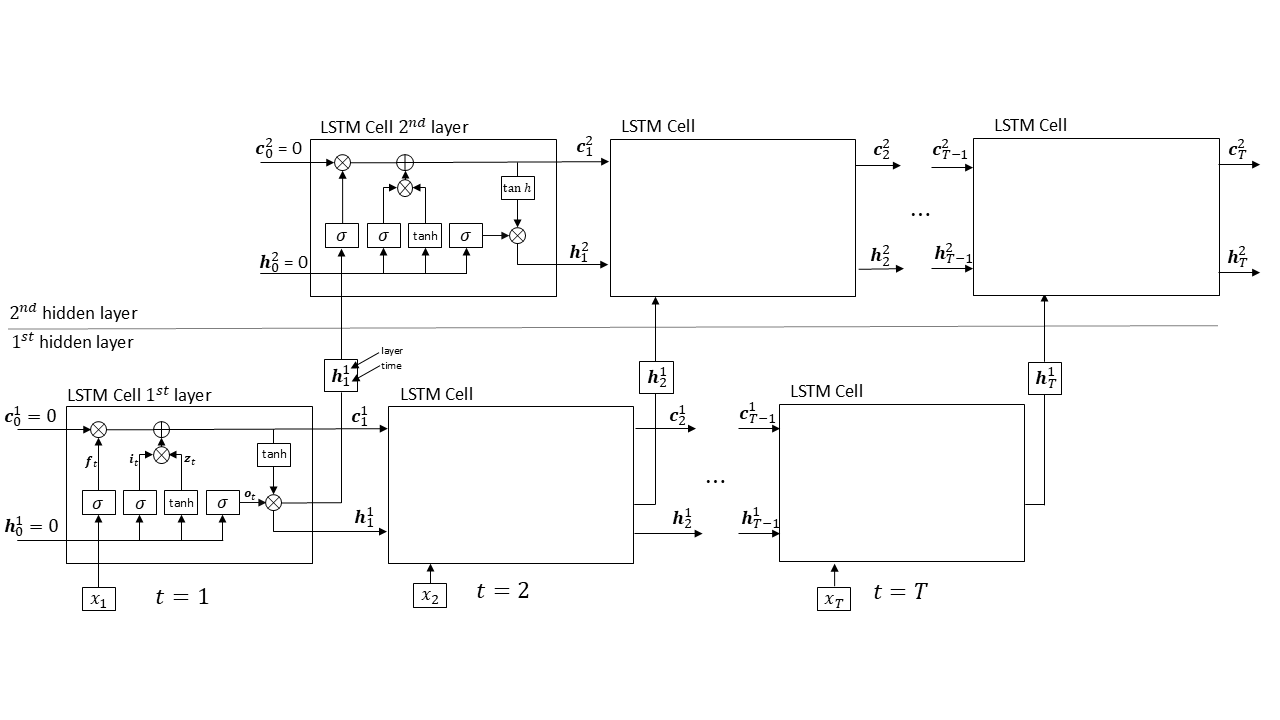}
\caption{\label{fig:lstm2layers}LSTM Architecture with 2 hidden layers.}
\end{figure}

When multiple hidden layers are used, the hidden state of lower layers act as input to the higher layers. In the first layer, the hidden state $\mathbf{h}_{t}^{(1)}$ is calculated for every day in the 60-day sequence. This sequence of hidden states then serves as the input for the second layer. Formally, for layer l, the operation is: $$\mathbf{h}_{t}^{(l)}=\text{LSTM}^{(l)}(\mathbf{h}_{t}^{(l-1)},\mathbf{h}_{t-1}^{(l)})$$ where $\mathbf{h}_{t}^{(0)}$ is the raw input vector $\mathbf{x}_{t}$. A two-layered LSTM can be found in Figure \ref{fig:lstm2layers}.

While the LSTM processes the sequence day-by-day, our forecasting task requires a single prediction for $t+1$. I therefore utilize a many-to-one LSTM architecture. After the $60^{th}$ return of stock $s$ in the sequence is processed, the model ignores all previous hidden states and extracts the final hidden state of the top layer, $\mathbf{h}_{T}^s$. This vector of size $H$ (set to 25 in the following), which are the latent features, represents the compressed temporal summary of the past T (=60 in my case) days. This summary is mapped to the output classes through a fully connected linear layer: $$\mathbf{z}_s=\mathbf{W}_{h}\mathbf{h}_{T}^s+\mathbf{b}$$ The logit vector $\mathbf{z}_s\in\mathbb{R}^{2}$ for stock $s$ is then passed through a Softmax function to produce a probability distribution across the two classes: $$\hat{y}_{s,i} = \text{softmax}(\mathbf{z}_s)_i = \frac{e^{z_{s,i}}}{\sum_{j=1}^{2} e^{z_{s,j}}}, \quad i \in \{0, 1\}$$

Since the model performs a binary classification, I use Cross Entropy as the objective function. This function penalizes the model based on how far the predicted probability is from the actual label (0 or 1). If the model predicts a high probability for the correct class y, the loss is low; if it predicts a high probability for the incorrect class, the loss increases exponentially. For a single observation, the loss L is calculated as: $$ \mathcal{L}_{s}=-\sum_{i\in\{0,1\}} y_{s,i}\,\log(\hat{y}_{s,i})=-[y_{s,0}\log(\hat{y}_{s,0})+y_{s,1}\log(\hat{y}_{s,1})] $$

The optimization of LSTM parameters requires a specialized form of gradient descent known as Backpropagation Through Time (BPTT). Unlike standard backpropagation used in feed-forward networks, BPTT must account for the temporal dependencies by unrolling the LSTM through each of the 60 time steps. During the forward pass, the model processes the sequence $\{x_{1},\dots,x_{60}\}$ and stores the internal gates $\mathbf{f}_{t}$, $\mathbf{i}_{t}, \mathbf{o}_{t}$ and states $\mathbf{c}_{t}$, $\mathbf{h}_{t}$ for every step. When the final loss $\mathcal{L}$ is computed at t=60, the gradient is propagated backwards not only through the layers but also backwards from t=60 to t=1. This allows the network to learn which specific historical return patterns at the beginning of the window are most predictive of the target at the end. To stabilize this process, I utilize the Adam optimizer, which computes individual adaptive learning rates for different parameters.

\subsection{Extension to the Canonical LSTM}
Three main extensions are made to the canonical LSTM: 1) the inclusion of additional macro-financial covariates; 2) accounting for cross-sectional heterogeneity by including sector information and 3) various forms of regularization.

\subsubsection{Additional Covariates}
The rationale for including macro-financial covariates is that they might help the model perform better during market regime changes which is hard to learn from the time-series of asset returns alone \parencite{guEmpiricalAssetPricing2020}. The way these covariates are integrated into the LSTM is quite simple: Next to passing the sequence of 60 lagged stock returns, I also pass the sequences of 60 lagged values of the selected covariates. Therefore, $\mathbf{x}_{t}$ from Figure \ref{fig:lstmCell} now contains not only the past return for time $t$ but also all other covariate values at time $t$. A detailed overview of covariates and their pre-processing can be found in section 3.1. 

\subsubsection{Cross-Sectional Heterogeneity}

A limitation of the LSTM approach in \cite{fischerDeepLearningLong2018} lies in using the exact same parameters of the neural network for each stock. The model treats every stock as if it belongs to the same underlying distribution. However, stocks in different sectors (e.g., Technology vs. Utilities) can exhibit distinct lead-lag effects and sensitivities to market shocks \parencite{houIndustryInformationDiffusion2007, moskowitzIndustriesExplainMomentum1999}. To capture this heterogeneity, I extend the architecture by incorporating a learnable sector embedding for TRBC sectors.

Let each stock $s$ be associated with a sector $c\in\{1,\dots,C\}$, where $C$ is the total number of unique sectors. The embedding operation acts as a differentiable lookup table. Mathematically, the retrieval of a sector embedding $\mathbf{e}_{c}$ is equivalent to a matrix multiplication:$$\mathbf{e}_{c}=\mathbf{W}_{emb}^{\top}\mathbf{v}_{c}$$ where $\mathbf{v}_{c}$ is a one-hot encoded vector representing the sector $c$ and $\mathbf{W}_{emb}\in\mathbb{R}^{C\times D}$ is the embedding weight matrix and D is the embedding dimension. The weight matrix $\mathbf{W}_{emb}$ contains the learnable parameters for all C sectors:
$$\mathbf{W_{emb}} = \begin{bmatrix}
w_{1,1} & w_{1,2} & \cdots & w_{1,D} \\
w_{2,1} & w_{2,2} & \cdots & w_{2,D} \\
\vdots  & \vdots  & \ddots & \vdots  \\
w_{C,1} & w_{C,2} & \cdots & w_{C,D}
\end{bmatrix}$$ 

Crucially, although the vector $v_{c}$ contains only 0s and a single 1, the values within $\mathbf{W}_{emb}$ are continuous real numbers. During the training phase, the model computes the gradient of the loss function $\mathcal{L}$ with respect to the embedding weights $\frac{\partial\mathcal{L}}{\partial\mathbf{W}_{emb}}$. Because $\mathbf{e}_{s}$ is concatenated with the LSTM output before the final classification, the gradients flow back from the loss through the fully connected layer and directly into the specific row of $\mathbf{W}_{emb}$ corresponding to the stock's sector. This allows the model to re-position sectors in the D-dimensional latent space. Over time, the model clusters sectors that respond similarly to certain temporal patterns (see later discussion for Figure \ref{fig:sector07weights}).

The LSTM processes the return sequence $\mathbf{X}_{t}^{s}=\{\tilde{R}_{t-m}^{s}\}_{m=0}^{59}$ to produce a final hidden state which represents the temporal features of the stocks as in the canonical version. The difference to the canonical version arises in the last layer, as the fully connected layer now has $H$ + D input nodes (H hidden units and D for embedding dimension) to get the logits for the two classes. Formally, the final decision is based on a logit vector $\mathbf{z}_{s}\in\mathbb{R}^{2}$, representing the two classes (above or below median return). It is the sum of the temporal signal, the sector loading, and the class bias:

$$\mathbf{z}_{s}=\underbrace{\mathbf{W}_{h}\mathbf{h}_{T}^{s}}_{\text{Temporal Signal}}+\underbrace{\mathbf{W}_{e}\mathbf{e}_{c}}_{\text{Sector Loading}}+\mathbf{b}$$ where $\mathbf{W}_{h}$ and $\mathbf{W}_{e}$ are the weight matrices for the temporal signal and the sector embeddings respectively. The temporal signal captures the momentum patterns extracted from the 60-day lag sequence of returns. The Sector Loading is a cross-sectional intercept which shifts the probability of a stock being classified as above median based on its industry group. For example, during a market-wide rally in Technology, the weights $\mathbf{W}_{e}$ and embeddings $\mathbf{e}_{s}$ will combine to produce a positive offset for Tech stocks, lowering the threshold required for the LSTM signal to trigger a Class 1 prediction. While the LSTM weights are shared across all stocks to learn general temporal patterns, the embedding layer learns a low-dimensional representation of sector-level structures. This facilitates the classifier in learning how temporal return patterns might imply different future outcomes depending on the sector context.

\begin{figure}
\centering
\includegraphics[width=0.4\linewidth]{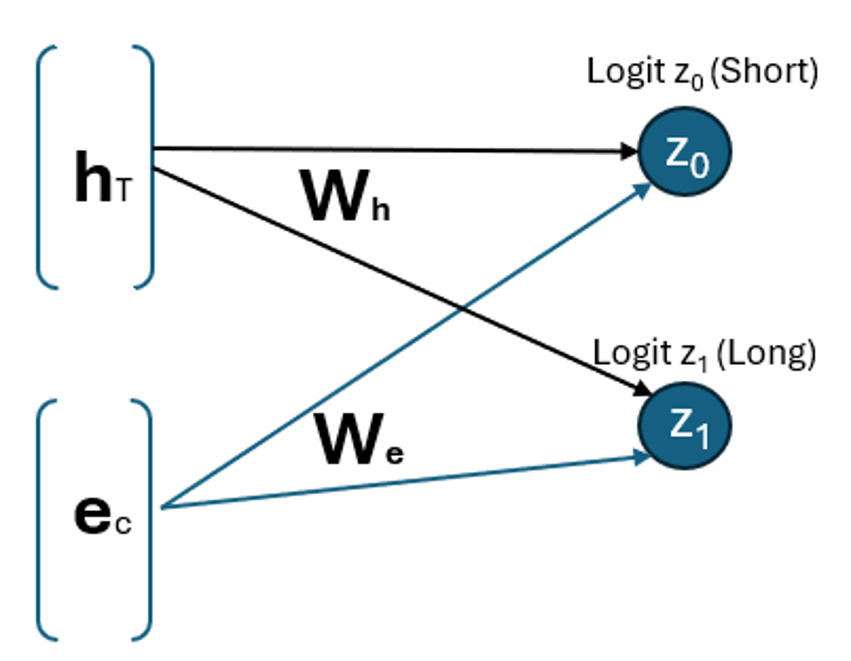}
\caption{\label{fig:sectorLogits}Incorporating sector embeddings into the output layer.}
\end{figure}

\subsubsection{Regularization}
A number of regularization techniques are applied, namely label smoothing, gradient clipping, weight decay, dropout and early stopping, to reduce the risk of overfitting. The last two regularization measures were already applied in \cite{fischerDeepLearningLong2018}. 

Instead of using binary targets (0 or 1), I use label smoothing to soften the targets. Label smoothing redistributes the confidence $ \alpha$  across all $K$ possible classes. The formula for the smoothed target $y_{s}^{LS}$ for a true class $y$ is \parencite{mullerWhenDoesLabel2019}:$$y_{s}^{LS}=y\cdot(1-\alpha)+\frac{\alpha}{K}$$ In my case, K=2 and $\alpha=0.1$ which results in the smoothed targets 0.05 and 0.95. Because of the low signal-to-noise ratio in finance, the labels are noisy and it is unrealistic for the model to have a high confidence for any stock under- or outperforming the market. 

Gradient Clipping helps to stabilize the training of the LSTM and prevent the exploding gradient problem common in recurrent architectures \parencite{zhangWHYGRADIENTCLIPPING2020}. Therefore, I clip the global norm of the gradients at $2.0$, similar to \textcite{liBinaryValuedGatesRobust2018}.

Weight decay helps in keeping the weights of the LSTM small. If the weights are large, this typically means that the network is memorizing training data or strongly reacts to changes in data. Weight decay subtracts a small fraction of the current weight at each iteration: $$w_{t+1} = w_t - \eta \nabla f(w_t) - \eta \lambda w_t$$ where $w_t$ is the current weight, $\eta$ is the learning rate, $\nabla f(w_t)$ is the gradient of the loss function and $\lambda$ is the weight decay coefficient. This $\lambda$ is set to 1e-4 in my application.

Early Stopping is used to stop model training when it is obvious that the model is overfitting \parencite{bai2021understanding}. For that reason, the validation loss is monitored with a patience of 20 epochs. Training is terminated if no improvement in the validation loss is observed for 20 epochs, and the weights from the best-performing epoch are restored for the out-of-sample prediction phase.

Furthermore, there are two distinct dropout mechanisms applied: 1) Input Dropout: I apply a Bernoulli-distributed mask to the input vector at each of the 60 timesteps. Each input variable is dropped independently with a probability p. This approach is particularly critical for the covariate model. Without the input dropout, the model exhibited severe overfitting, characterized by a diverging validation loss despite a rapidly decreasing training loss. 2) Recurrent Dropout: To regularize the temporal dependencies, dropout is applied to the hidden state transitions. For each timestep, components of the 25-dimensional hidden state vector are randomly zeroed, preventing the model from over-relying on specific memory cells.

\subsection{Benchmark Models}
For benchmarking the LSTM models, a random forest model is used. Random forests are shown to be the best-performing individual method in \textcite{kraussDeepNeuralNetworks2017}. Furthermore, it is a well-established, low-maintenance model that reliably delivers strong out-of-sample performance. Random Forests were introduced by \textcite{ho1995random} and formalized by \textcite{breiman2001random}. They combine an ensemble of deep decision trees, each trained on a bootstrap sample of the data. Two mechanisms reduce variance: bagging, which trains each tree on a distinct bootstrap sample, and random feature selection, which restricts each split to a random subset of $m$ out of $p$ features, decorrelating the trees. Classification is performed by majority vote across all $B$ trees.
The input must be adjusted as these methods do not have memory and cannot process the time series of variables. Hence, the past returns must be included as single variables. Similar to \cite{fischerDeepLearningLong2018}, cumulative returns $R_{t}^{m,s}$ with $m\in\{\{1,2,...,20\}\cup\{40,60\}\}$ are the features and class labels based on the return $R_{t+1}^{s}$ are the targets. I follow standard hyperparameter choices: B=500 trees, maximum depth J=10, and $m=\sqrt{p}$ for feature subsampling.

As a market benchmark, I use the value-weighted US market return from the Kenneth R. French Data Library\footnote{\url{https://mba.tuck.dartmouth.edu/pages/faculty/ken.french/data_library.html}}, constructed as the sum of the market excess return factor and the risk-free rate. This market benchmark covers all NYSE, AMEX, and NASDAQ stocks and is available at daily frequency.

\section{Data}
\subsection{Data Source}

To ensure our empirical analysis is free from survivorship bias, I reconstructed the S\&P 500 index composition using monthly constituent lists from Refinitive (formerly Thomson Reuters). The sample covers the period from January 1995 to December 2024. The lists were converted into a binary matrix that identifies whether a stock was part of the index in any given subsequent month, allowing us to replicate the index’s historical membership accurately. Subsequently, I retrieved daily total return indices for every stock that appeared in the index during this timeframe. These indices are adjusted for dividends, stock splits, and corporate actions, providing a precise foundation for calculating investment returns. Table \ref{tab:sector_returns} contains yearly average summary statistics by sector, depicting the mean annual return the standard deviation of the annual return, the minimum and maximum annual return, the percentage of positive returns during the 30 years of data availability and the average number of stocks in these sectors. There are 11 sectors based on TRBC. 

\begin{table}[htbp]
\centering
\caption{Average yearly summary statistics for S\&P 500 constituents from January 1995 until December 2024, split by industry. They are based on equal-weighted portfolios per industry as defined by TRBC, formed on a monthly basis, and restricted to index constituency of the S\&P500. Returns and the standard deviation are denoted in percent.}
\label{tab:sector_returns}
\resizebox{\textwidth}{!}{%
\begin{tabular}{lcccccc}
\hline
Sector & Mean Return & Std.\ Dev. & Min Return & Max Return & Pct.\ Pos.\ Years & Avg.\ Num.\ Stocks \\
\hline
Technology                & 15.6 & 28.2  & $-44.2$ & 71.6 & 73.3 & 77.3 \\
Healthcare                & 13.6 & 18.8  & $-26.7$ & 53.1 & 76.7 & 51.4 \\
Industrials               & 11.9 & 17.0  & $-36.9$ & 38.2 & 76.7 & 66.6 \\
Energy                    & 11.4 & 27.7  & $-49.8$ & 51.7 & 63.3 & 32.8 \\
Financials                & 11.0 & 22.1  & $-53.3$ & 48.0 & 66.7 & 70.7 \\
Consumer Cyclicals        & 10.3 & 20.7  & $-41.9$ & 67.5 & 73.3 & 74.9 \\
Basic Materials           &  9.3 & 18.5  & $-35.3$ & 52.8 & 66.7 & 31.8 \\
Consumer Non-Cyclicals    &  9.1 & 12.3  & $-22.9$ & 31.3 & 80.0 & 45.4 \\
Educational Services      &  8.7 & 37.3  & $-35.5$ &  8.0 & 50.0 &  1.8 \\
Real Estate               &  8.5 & 19.8  & $-48.6$ & 45.2 & 70.0 & 16.9 \\
Utilities                 &  6.8 & 16.1  & $-32.1$ & 40.8 & 70.0 & 29.5 \\
\hline
\end{tabular}%
}
\end{table}

\subsection{Model Covariates and Data Preprocessing}

\begin{figure}
\centering
\includegraphics[width=1\linewidth]{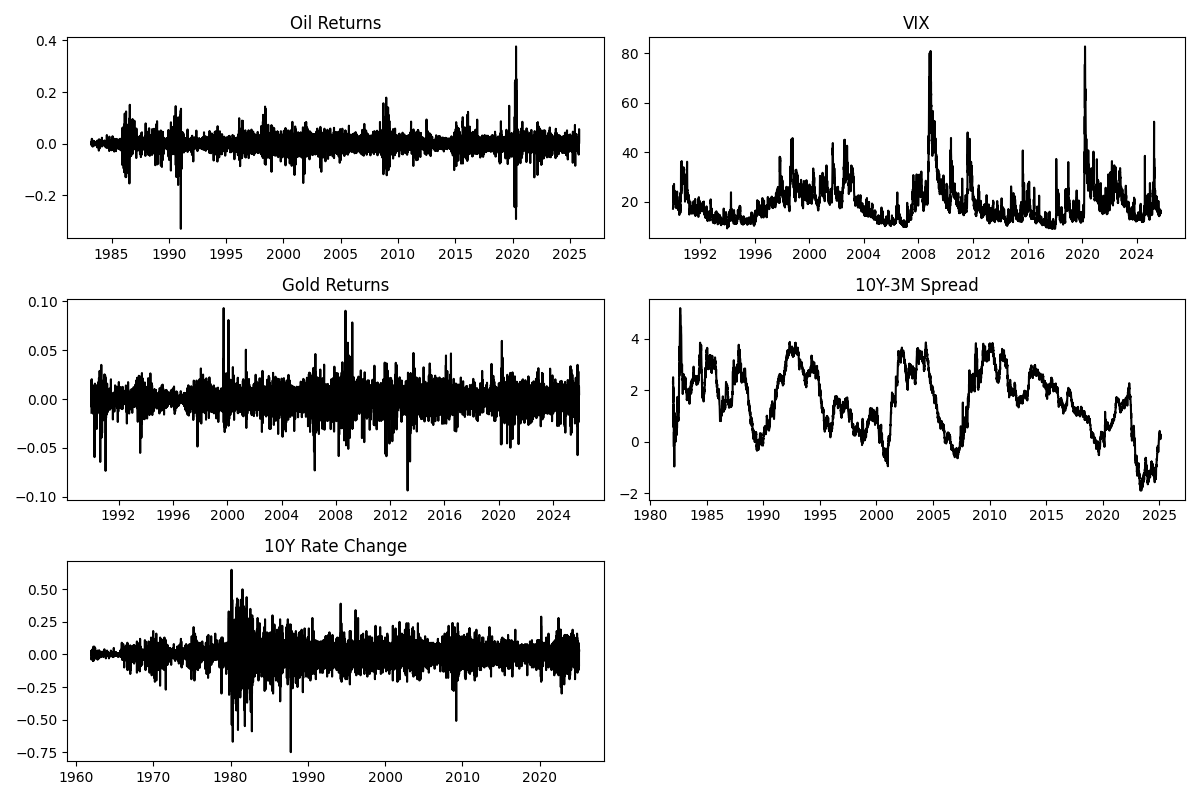}
\caption{\label{fig:covariates}Time Series for Covariates, Source: Bloomberg and Thomson Reuters}
\end{figure}

For one architectural extension of the LSTM model, I incorporate several macro-financial covariates consisting of oil returns, the CBOE Volatility index (VIX), gold futures, 10-year treasury minus 3-month treasury the and the treasury constant 10-year maturity. These variables capture systematic risk, liquidity conditions, and industrial demand, providing the model with broader economic context beyond historical price action. Other covariates were considered but not included due to certain shortcomings. For example, the $tedrate$ was discontinued on January 21, 2022 which renders it unusable for my prediction task over the whole forecasting period. Another potential variable, the spread between baa and aaa bonds, showed not enough variation to exploit for prediction purposes. The following paragraph motivates the use of these 5 variables and references papers that showed successful results incorporating these variables into stock market forecasting models.

Following \textcite{narayanHasOilPrice2015} and \textcite{elyasianiOilPriceShocks2011}, oil returns are included as a proxy for production costs. Fluctuations in energy prices create a dual effect, burdening energy-intensive industries while simultaneously benefiting petroleum producers. Market-wide risk and uncertainty are accounted for through the VIX, which serves as a benchmark for expected volatility and risk premiums in the S\&P 500 \parencite{bekaertVIXVariancePremium2014}, and gold futures, which function as a classic Safe Haven asset during periods of economic distress \parencite{huangForecastingTaiwanStock2023, beckmannGoldPriceDynamics2019}. The model further integrates the term structure of interest rates to capture shifting expectations regarding growth and monetary policy. The 10-year to 3-month Treasury spread is utilized as a primary recession indicator, as an inverted yield curve typically signals extreme systemic risk that reshapes sector-specific performance \textcite{estrellaYieldCurvePredictor1996}. Complementing this, the 10-year Treasury yield itself tracks the broader cost of capital. As noted by \textcite{chordiaEmpiricalAnalysisStock2005}, rising long-term rates tend to disproportionately impact the valuation of long-duration assets, such as high-growth technology stocks, while providing potential tailwinds for the financial sector through improved interest margins.

\bigskip
\textbf{Data Preprocessing}
\newline
To ensure numerical stability during the LSTM’s gradient descent, I address data gaps and extreme outliers:
There was an oil price anomaly on April 20, 2020, when WTI crude futures plummeted to negative values \parencite{frenchCrudeOilPrices2021}. To prevent this extreme outlier from distorting the model’s learning, I applied linear interpolation, resulting in a corrected value of 14.14 for that period. Six missing observations in the VIX series occurred between 1997 and 2021. Based on the dates, there was no specific pattern visible for the missing values which implies the data points were missing completely at random. Therefore, I also linearly interpolated the missing values to maintain a continuous time series for the prediction models. 

Furthermore, standardization is performed for each study period using only the training data to calculate mean and standard deviation. The standardization is applied to all values, including the test set, similar to the stock returns discussed in section 2.1:

$$\tilde{X}_{t}=\frac{X_{t}-\mu_{train}}{\sigma_{train}}$$

\section{Results}

This section provides an overview of the results of the extended LSTM neural networks against the benchmark models, validates the long-term performance of the LSTM models and takes a closer look at the interpretability of the learned sector embeddings and its ties to the momentum literature.

\subsection{Performance against Benchmarks}

\begin{figure}
\centering
\includegraphics[width=0.9\linewidth]{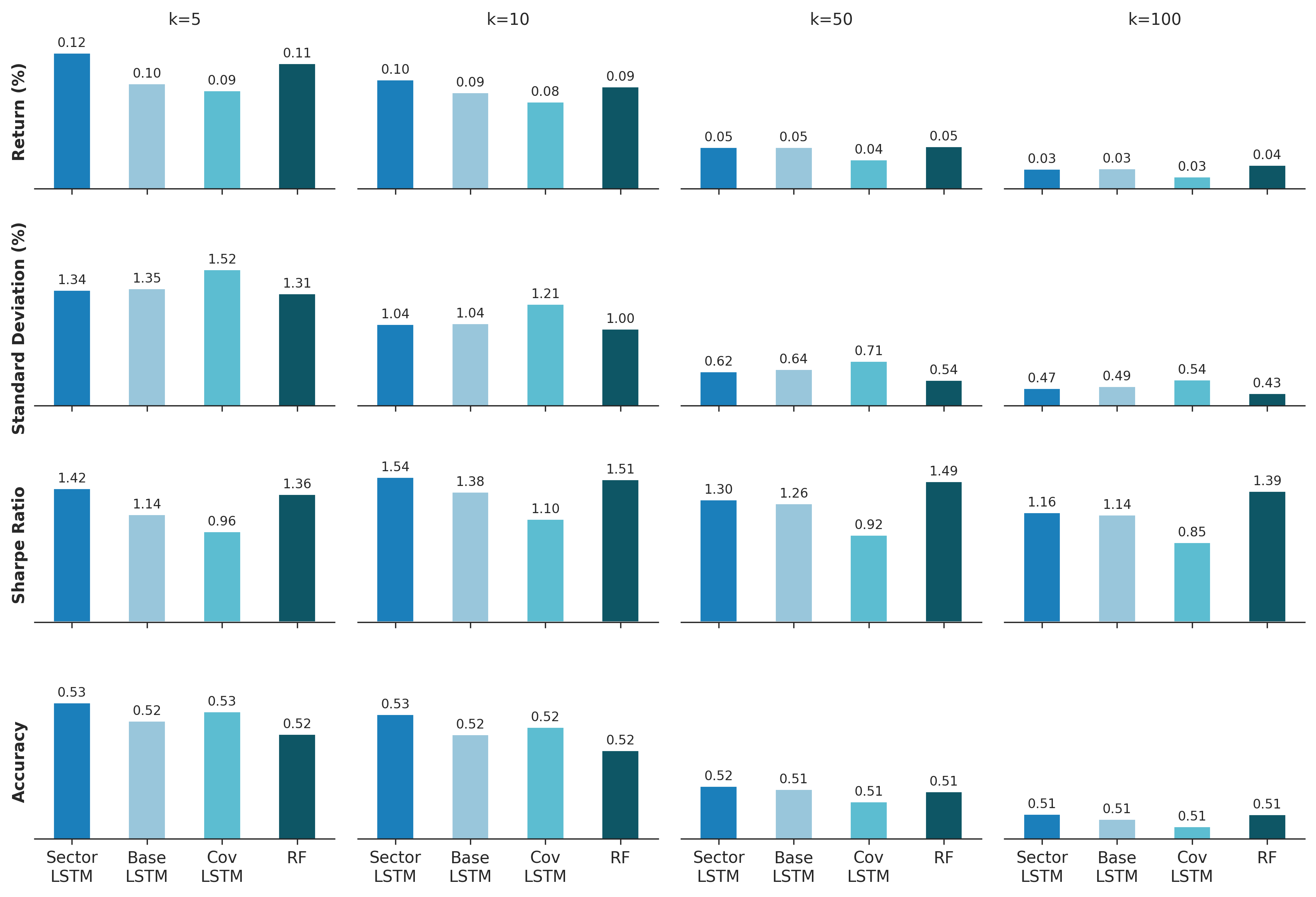}
\caption{\label{fig:performance}Daily performance characteristics for long-short portfolios of different sizes: mean return (excluding transaction costs), standard deviation, annualized Sharpe ratio (excluding transaction costs), and accuracy from January 1995 to December 2024}
\end{figure}

To compare the main performance characteristics, I evaluate the sector LSTM and covariate LSTM to the base LSTM and the random forest model in Figure \ref{fig:performance} for different portfolio sizes $k$. In this case, $k$ states the number of long and short positions, meaning the portfolio has in total $2k$ positions. The four metrics are daily return, standard deviation of daily returns, accuracy (all in \%) and annualized Sharpe ratio. The Sharpe ratio is calculated by dividing the mean daily long-short portfolio return by the standard deviation of those daily returns across the full sample period, annualized by multiplying by $\sqrt{252}$. Accuracy is the average fraction of stocks in the long and short positions whose direction was correctly predicted, averaged across all trading days.
What can be seen from Figure \ref{fig:performance} is that the Sector LSTM has the best performance for Daily return and accuracy. The RF has generally a low standard deviation and is close to the sector LSTM for the Sharpe Ratio or even better for larger $k$. The base LSTM performs slightly worse than the sector LSTM in all metrics. In contrast, the covariate LSTM has the lowest average daily returns, the highest standard deviation and lowest Sharpe ratio. Hence, it's performance is the worst of the considered models.

To formally assess whether the differences in predictive accuracy across models are statistically significant, I apply the Diebold-Mariano (DM) test following the construction in \textcite{fischerDeepLearningLong2018}. The null hypothesis is that two competing models have equal predictive accuracy; a rejection in favor of the row model indicates statistically superior forecasting performance. The results are reported in Table \ref{tab:dm_test}.

\begin{table}[h]
\centering
\caption{Diebold-Mariano-Test $p$-values}
\label{tab:dm_test}
\begin{threeparttable}
\begin{tabular}{lcccc}
\toprule
 & Sector LSTM & Cov LSTM & Base LSTM & RF \\
\midrule
Sector LSTM & $-$        & 0.1891    & 0.0915*   & 0.0095*** \\
Cov LSTM    & 0.8109     & $-$       & 0.2849    & 0.0441**  \\
Base LSTM   & 0.9085     & 0.7151    & $-$       & 0.1151    \\
RF         & 0.9905     & 0.9559    & 0.8849    & $-$       \\
\bottomrule
\end{tabular}
\begin{tablenotes}
\small
\item \textit{Note:} * $p < 0.1$, ** $p < 0.05$, *** $p < 0.01$. Null hypothesis: equal predictive accuracy. Rejection in favor of the row model indicates superior forecasting performance.
\end{tablenotes}
\end{threeparttable}
\end{table}

 The sector LSTM significantly outperforms the Random Forest at the 1\% level (p = 0.0095), providing strong evidence that the architectural extensions to the canonical LSTM yield genuine predictive gains over the tree-based benchmark. Against the base LSTM, the sector LSTM achieves marginal significance at the 10\% level (p = 0.092). This weaker result is economically interpretable: the sector embedding enters the model as a static cross-sectional intercept rather than a time-varying input. In essence, the two models share identical temporal processing of the return sequence and differ only in the assignment of cross-sectional weights at the classification stage. The incremental signal from sector information is therefore modest in magnitude, and the DM-Test reflects this accordingly. Notably, the base LSTM does not significantly outperform the Random Forest (p = 0.115), suggesting that the temporal structure captured by the LSTM alone, without sector information, does not translate into a statistically distinguishable accuracy advantage over the RF benchmark model. The covariate LSTM is not outperforming the canonical LSTM indicating that additional macro-financial covariates do not help the model to predict the direction of stocks in the S\&P 500. Against the RF, the covariate LSTM achieves significance at the 5\% level. It is important to note that the DM-Test only test for accuracy and does not include any risk or return metrics. These will be tested in the next step. 

\begin{table}[htbp]
\centering
\caption{Ledoit--Wolf Test $p$-values for Differences in Sharpe Ratios}
\label{tab:lw_sharpe}
\begin{tabular}{lccccc}
\toprule
 & Sector LSTM & Cov LSTM & Base LSTM & RF & Market \\
\midrule
Sector LSTM & --     & 0.009*** & 0.045**  & 0.204    & 0.177    \\
Cov LSTM    & 0.991  & --       & 0.795    & 0.917    & 0.769    \\
Base LSTM   & 0.955  & 0.205    & --       & 0.754    & 0.560    \\
RF          & 0.796  & 0.083*   & 0.246    & --       & 0.371    \\
Market      & 0.823  & 0.231    & 0.440    & 0.629    & --       \\
\bottomrule
\end{tabular}

\vspace{0.5em}
\begin{minipage}{0.92\textwidth}
\footnotesize
\textit{Note:} * $p<0.10$, ** $p<0.05$, *** $p<0.01$. One-sided test; the null
hypothesis is equal Sharpe ratios, and rejection in favor of the row model
indicates a significantly higher Sharpe ratio. Annualized Sharpe ratios are
computed on $N=6{,}793$ daily long-short returns. Standard errors are obtained
from a studentized circular block bootstrap (Ledoit and Wolf, 2008) with a block
length of 19.
\end{minipage}
\end{table}

To test differences in Sharpe ratios, I apply a one-sided circular block bootstrap following \textcite{ledoitRobustPerformanceHypothesis}, who develop a Sharpe Ratio test robust to the non-normal, heteroskedastic, and autocorrelated returns typical of financial data. The results can be found in Table \ref{tab:lw_sharpe}. For each pair I compute the difference in annualized Sharpe ratios on the common sample of $N = 6,793$ daily long-short excess returns and test $H_0: SR_i = SR_j$ against $SR_i > SR_j$; rejection in favor of the row strategy indicates a significantly higher Sharpe ratio. Under the null, both series are resampled jointly with a block length of 19 days, preserving each series' serial dependence and the cross-correlation between strategies. The block length is set to $N^{1/3} \approx 19$, following the rate-optimal rule of \textcite{hallBlockingRulesBootstrap1995}. The bootstrap iterations are set to 1,000.

The sector LSTM has a significantly higher Sharpe ratio than the base LSTM (0.69 vs. 0.39, p = 0.045) and the covariate LSTM (0.69 vs. 0.27, p = 0.009). However, the sector LSTM is not performing significantly better than the random forest (0.69 vs. 0.53, p = 0.204). The covariate LSTM is dominated, with a significantly lower Sharpe ratio than both the sector LSTM (p = 0.009) and the random forest (p = 0.083), giving formal support to the conclusion that macro-financial covariates are not beneficial for the LSTM in this application. Overall, an interesting insight is that the covariate LSTM is performing poorly on the risk-return metrics, but is performing well at accuracy. In the following parts of the results section, I will focus on the extension regarding the sector information. 

To evaluate the practical viability of the LSTM strategy, transaction costs must be considered as the long and short positions are decided upon every day and therefore turnover is high. Generally, the long-short trading strategies are primarily used by institutional investors and hedge funds. \textcite{frazzini2012tradingcosts} measured trading costs from an institutional investor with real-world data and find trading costs that are roughly an order of magnitude smaller than claimed in the literature. Additionally, my strategy only considers large-cap liquid U.S. stocks which have the lowest bid-ask spreads \parencite{asnessSizeMattersIf2018}. Assuming a trader uses only a fraction of daily volume to build the positions, assuming 0.01\% to 0.03\% of total daily trading volume, \textcite{frazzini2012tradingcosts} show that the median market impact is negligible. Therefore, I use tradings costs of 2 basis points per half-turn that is used in previous studies \parencite[e.g.][]{jhaTimingEquityQuant2016}. These trading costs represent an environment for modern institutional traders dealing in highly liquid S\&P 500 stocks, where minimal commissions and low bid-ask spreads lead to a small total cost per trade. The 2 basis points are applied twice for every stock, when building the long/short position and when closing the position. If a stock is in the long leg of the portfolio e.g. for 3 consecutive days then trading costs are only applied on the first day when buying the stock and on the third day when selling the stock.
\newline

\begin{figure}
\centering
\includegraphics[width=0.8\linewidth]{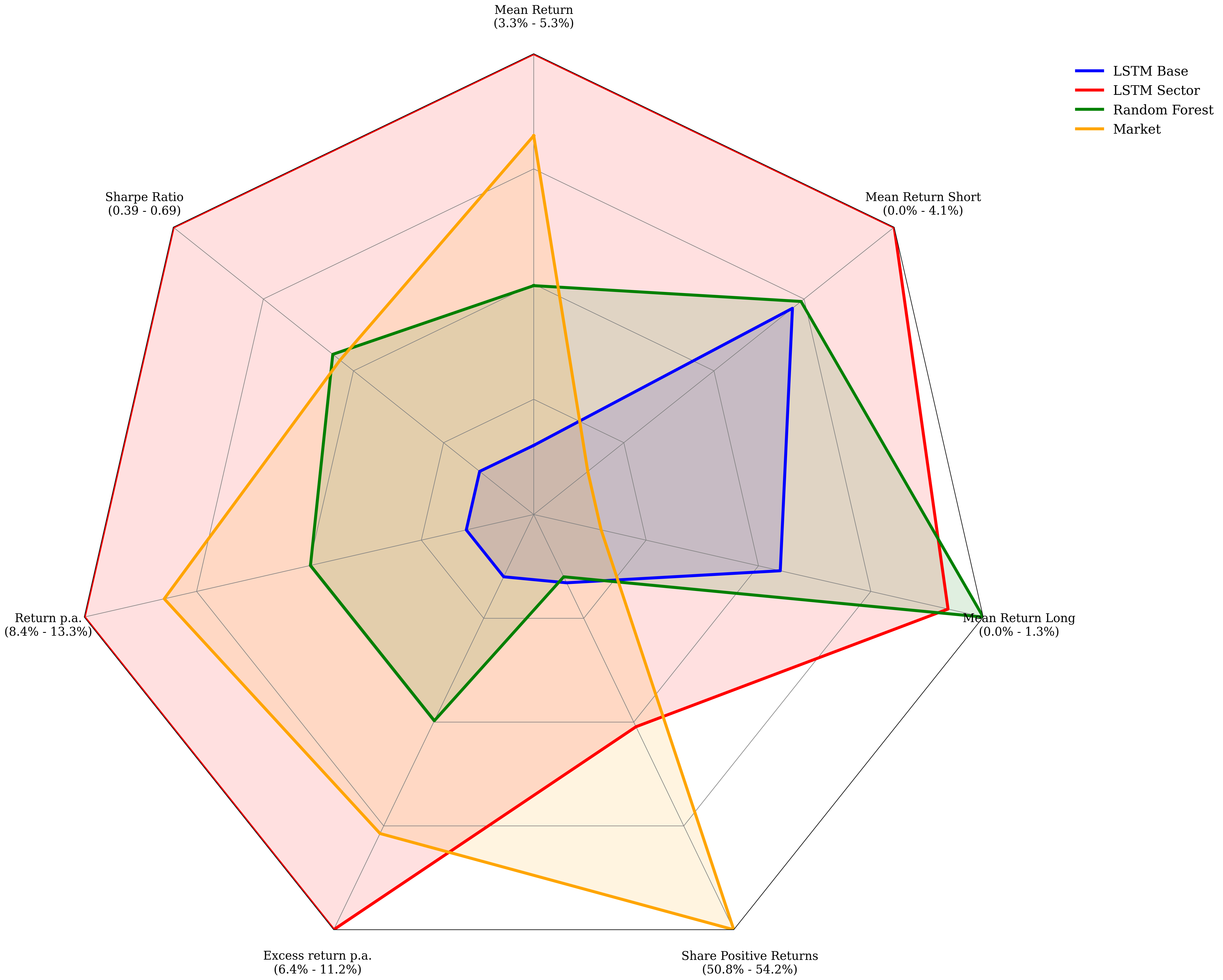}
\caption{\label{fig:returnCharacteristics}Return characteristics of LSTM sector model against three benchmarks (after transaction costs). A larger surface area is more favourable for the model.}
\end{figure}

\begin{figure}
\centering
\includegraphics[width=0.8\linewidth]{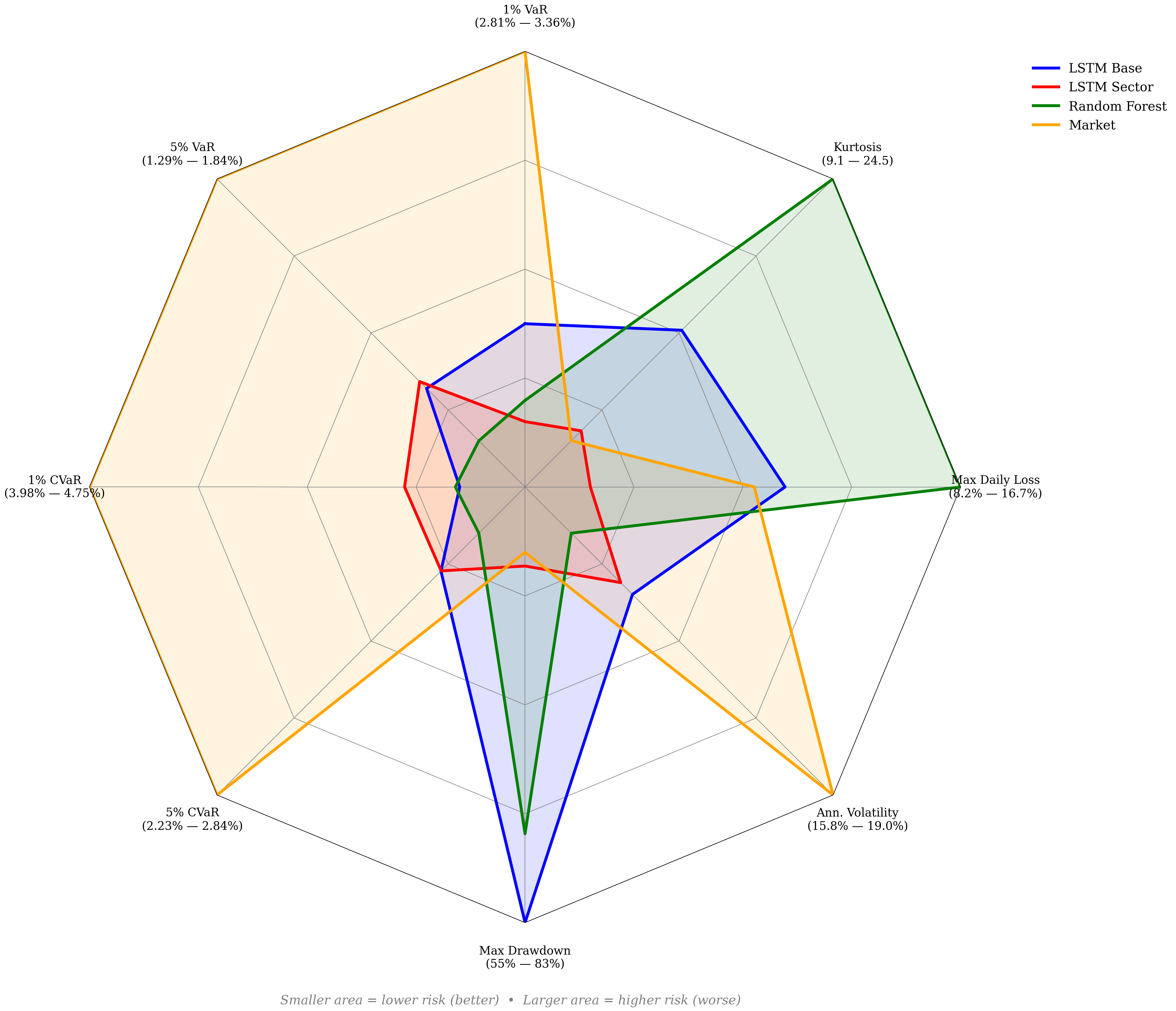}
\caption{\label{fig:riskCharacteristics}Risk characteristics of LSTM sector model against three benchmarks (after transaction costs). A smaller surface area is more favourable for the model.}
\end{figure}

\noindent \textbf{Return Characteristics}
\newline
The radar chart in Figure \ref{fig:returnCharacteristics} illustrates return characteristics of the LSTM sector model against three benchmarks: base LSTM model, random forest and the buy-and-hold-strategy for a market portfolio. A larger surface corresponds to more favorable return characteristics. For all models except the buy-and-hold strategy, transaction costs of 2 basis points per half turn were applied. The return characteristics include 1) daily variables: daily mean return, daily mean return for short leg, daily mean return for long leg and share of positive returns and 2) yearly variables: excess return, overall return and Sharpe Ratio. After trading costs, the base LSTM model and the random forest have lower average returns than the buy-and-hold strategy. The sector LSTM performs better than all the benchmarks in each category except share of positive returns (less than the buy-and-hold strategy) and the mean return long, where the performance is close to the random forest model. These results demonstrate that the sector LSTM is the dominant model across the return metrics, although long-short portfolios are generally not exposed to market risk and therefore capture no $\beta$. One detail to note, is that the short and long return of the market where set to 0. The exact numerical values for each model can be found in table \ref{tab:return-characteristics} in the appendix.
\newline

\noindent\textbf{Risk Characteristics}
\newline
To complement the return-based evaluation, Figure \ref{fig:riskCharacteristics} presents eight risk metrics for each model, where a smaller enclosed area indicates more favorable risk characteristics. The metrics capture tail risk, distributional properties, volatility, and drawdown behavior of the long-short portfolios.

The Value-at-Risk at the 1\% and 5\% confidence levels (1\% VaR, 5\% VaR) represent the maximum daily loss not exceeded on 99\% and 95\% of trading days respectively, providing a threshold-based measure of downside exposure. The Conditional Value-at-Risk at the same confidence levels (1\% CVaR, 5\% CVaR), also known as Expected Shortfall, goes one step further by measuring the expected loss conditional on returns falling beyond the VaR threshold. CVaR is therefore a more conservative and coherent risk measure that captures the severity of tail losses. See \textcite{yamaiValueatriskExpectedShortfall} for an overview of these risk metrics. In this analysis, CVaR is expressed as the average loss of the portfolio in percent. Annualized Volatility measures the standard deviation of daily returns scaled to an annual frequency, capturing overall return dispersion. Maximum Drawdown represents the largest peak-to-trough decline over the full sample period, reflecting the worst sustained loss an investor would have experienced. Finally, Kurtosis measures the fat-tailedness of the return distribution relative to a normal distribution, where higher values indicate a greater frequency of extreme return realizations.

\begin{figure}
\centering
\includegraphics[width=0.89\linewidth]{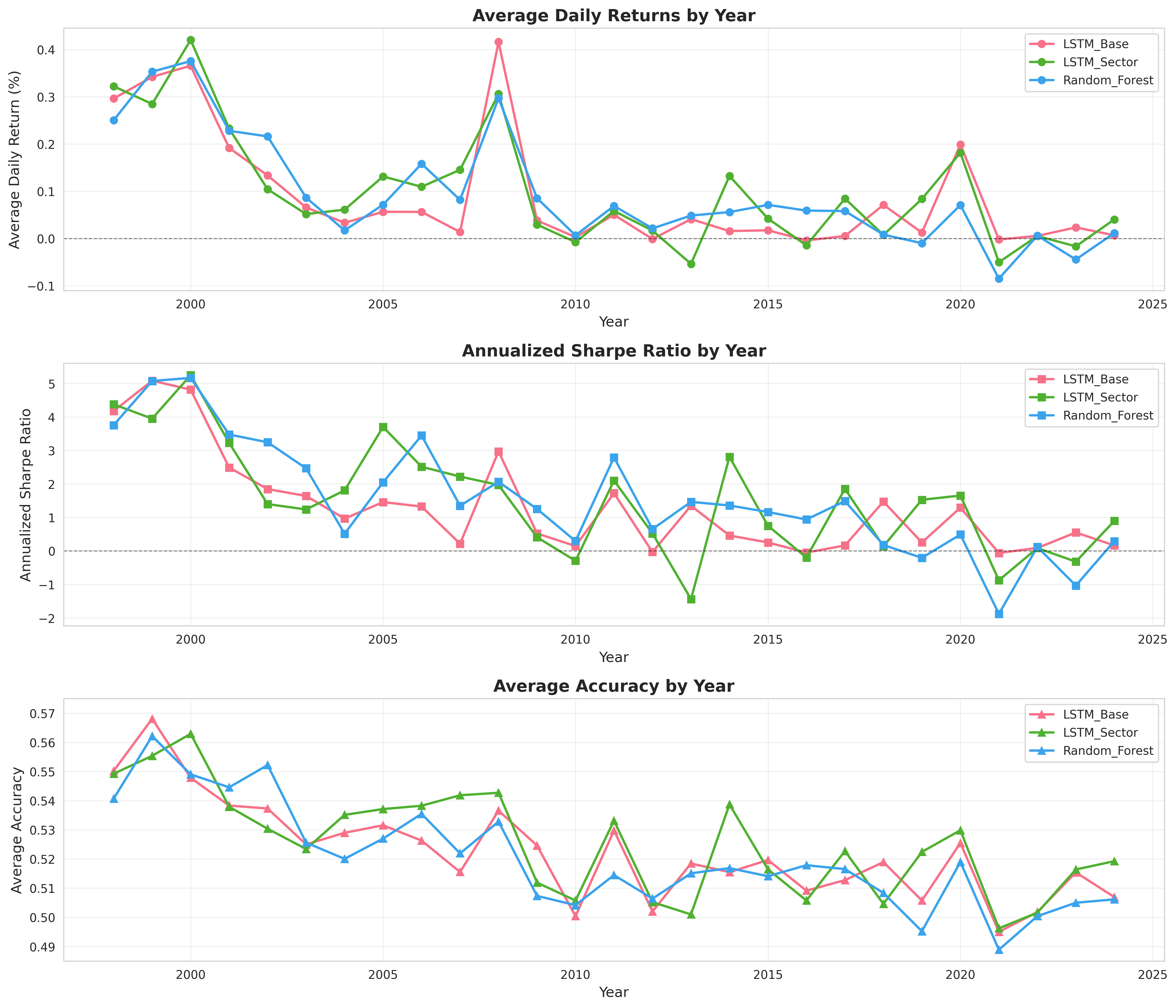}
\caption{\label{fig:longtermMetrics}Performance Metrics by Year}
\end{figure}

The market benchmark has considerably more exposure to systematic risk compared to all other models, which is expected given the nature of the long-short strategy. A market-neutral long-short portfolio has no net equity exposure and therefore does not earn the equity risk premium, but in exchange it avoids the market risk embedded in a buy-and-hold market position. The market's 1\% VaR and CVaR values of up to 3.36\% and 4.75\% respectively reflect the possibility of high losses on single days such as during the 2008 financial crisis and the COVID-19 shock of 2020. The long-short portfolios, by construction, are largely protected from these systemic events, which is precisely their appeal as a diversifying strategy.

Among the three active strategies, the sector LSTM exhibits the most favorable risk profile overall. It has the smallest surface area of all models and is at or close to the minimum values for every risk metric. Notably, the random forest displays markedly elevated kurtosis relative to both LSTM specifications, suggesting a heavier-tailed return distribution. The huge maximum daily loss of 16.7\% for random forest occurred on March 9, 2020 during the stock market crash triggered by COVID-19.

The base LSTM exhibits a deeper drawdown than the sector LSTM, pointing to extended periods of consecutive losses that the sector information helps to mitigate. The sector LSTM's shallower drawdown suggests that the sector-level bias in stock selection not only improves average returns but also provides a degree of stabilization during periods when the temporal signal from the LSTM sequence alone is insufficiently discriminative.

\subsection{Long-Term Validation of LSTM Performance}

The excessively high returns of the LSTM reported in \textcite{fischerDeepLearningLong2018} (average annualized return p.a. of $>200\%$ before transaction costs and $82\%$ after transaction costs) are not maintainable in the long run. These returns were driven by returns in the 1990's where in some years the Sharpe Ration was close to 20. While I can generally replicate the accuracy metrics, their daily returns differ by some margin from my results even for the years where our out-of-sample years overlap. 

Figure \ref{fig:longtermMetrics} displays the average daily returns, Sharpe ratio and accuracy, for the base and the sector LSTM as well as the random forest for each year of the forecasting time horizon. Although the sector LSTM slightly outperforms the other methods, the results seem robust as all models have similar patterns, e.g. decline from 2009 onwards and good metrics in 2008 and 2020.

\textbf{Average Daily Returns By Year}: All 3 models performs well in the years prior and shortly after the year 2000. Afterwards the performance declines noticeably except for two years. In times of economic uncertainty, namely 2008 during the financial crisis and 2020 in the midst of the COVID pandemic, the models show a high daily return. This finding is in line with studies \parencite{cleggPairsTradingPartial2018, jacobsDeterminantsPairsTrading2015} suggesting higher arbitrage opportunities during times of financial uncertainty. The base LSTM model even has its best year in terms of daily return in 2008. Generally, the base LSTM is the worst performing method in the majority of years with minor exceptions during the end of the study time frame. After 2008, the performance of all 3 models is consistently below 0.1\% average daily return, which confirms the result of \textcite{fischerDeepLearningLong2018} which find a deteriorating performance of LSTM models for statistical arbitrage on the S\&P500.

\textbf{Annualized Sharpe Ratio by Year}: Overall, the Sharpe Ratio is declining over the study periods. In the early study periods, all models had Sharpe Rations above 4. Until 2008, the Sharpe Ratios are declining but still stay approximately at 2 on average. After 2009, there are even some years with negative Sharpe Ratios. From 2018 until 2024, the random forest seems to perform especially poor, as its Sharpe Ratio is either marginally above 0 or negative. Similar to the daily returns graph above, the base LSTM seems to perform the worst overall with exception for the last few years of the study period.

\textbf{Average Accuracy by Year}: Here, the sector LSTM has a visible advantage with highest accuracy in the majority of the years. In the period from 2004 to 2009, the sector LSTM consistently has a higher accuracy then the other two models. The accuracy from all three models decline over the period of 27 years of out-of-sample forecasting. The random forest even has accuracy of less than $50\%$ in 2021. 

In conclusion, the strongly positive returns and high Sharpe Ratios only exist up until 2008. With the exception of 2020, there is no year afterwards, where all the models perform exceptionally well. This confirms 
\textcite{bogomolovPairsTradingBased2013} and \textcite{radProfitabilityPairsTrading2016} who  find deteriorating profits for quantitative trading strategies.

\subsection{Decomposing the Predictive Signal for Interpretability: Industry Momentum and Short-Term Reversal}

\textcite{fischerDeepLearningLong2018} already hypothesize that the base LSTM can learn sector information based on the past returns only. My approach using sector embeddings includes a more direct learning chance for the network by including the sector information in the final linear layer. This architecture avoids any type of vanishing/exploding gradient problem as the sector information, static in nature, are not included in the temporal information of LSTM network. The main assumption for a positive effect in the return metrics is that the sector performance in the three training years is similar to sector performance in the test year. 

A central critique of deep learning models in finance is their lack of interpretability, commonly referred to as the black-box problem. The architectural design of the Sector-LSTM, however, permits a decomposition of the predictive signal into two economically meaningful and theoretically grounded components. The first is a cross-sectional intercept derived from the learnable sector embeddings, which functions as an implicit industry momentum factor. The second is a temporal signal extracted from the 60-day return sequence by the LSTM, which captures short-term reversal patterns at the individual stock level. Crucially, both components emerge endogenously from gradient descent on the return prediction objective, rather than being imposed by the researcher. Their correspondence to well-documented factors in the empirical asset pricing literature thus constitutes evidence that the neural networks can recover known factors.

\subsubsection{Decomposing the Sector Contribution}

\begin{figure}
\centering
\includegraphics[width=0.4\linewidth]{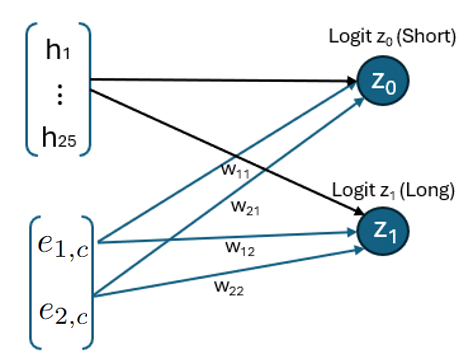}
\caption{\label{fig:finalLayer}Structure of the final linear layer.}
\end{figure}

To validate the economic intuition captured by the model, I decompose the final linear transformation layer to extract the contribution of the sector embeddings to the classification logits. Figure \ref{fig:finalLayer} shows the structure of the last hidden layer connecting to the logits before these are transformed into probabilities using the softmax function. This figure shows my empirical setup, e.g. I have 25 hidden units like \textcite{fischerDeepLearningLong2018} and a 2-dimensional vector for each sector. The model concatenates the LSTM hidden state $\mathbf{h}$ with the learned sector embedding $\mathbf{e_{c}}$. The analysis focuses on the weights $w_{ij}$ connecting the embedding dimensions to the output logits $(z_{0}$,$z_{1}$) to quantify a sector-specific bias. To check if the model favoured a sector to be classified into Class 1, the contribution of the sector embeddings to the logits must be isolated by computing the difference $\Delta_c=(e_{1,c}w_{12}+e_{2,c}w_{22})-(e_{1,c}w_{11}+e_{2,c}w_{21})$. If $\Delta_c>0$, the learned embedding for sector c inherently biases the model toward a classification into Class 1. This means that stocks in sector $c$ require a weaker temporal signal to be assigned to the long portfolio. Conversely, $\Delta_c<0$ introduces a systematic bias toward the short portfolio for that sector. This scalar quantity thus serves as an interpretable, model-derived measure of the sector-level return expectation learned during training. 

\subsubsection{Industry Momentum in the Sector Embeddings}

The sector contribution $\Delta_c$ relates closely to the momentum literature: In a 3-month to 1-year time horizon, \textcite{jegadeeshReturnsBuyingWinners1993} find that buying past winners and selling past losers can earn significant positive alphas. \textcite{moskowitzIndustriesExplainMomentum1999} show that momentum is induced by industries where high momentum industries perform better than low momentum industries six months after creating portfolios. The Sector-LSTM captures an analogous mechanism: during the three-year training window, the model learns which sectors have exhibited persistently strong or weak relative performance, and encodes this information as a directional bias in the embedding space. In the test year, which falls into the time horizon from \textcite{jegadeeshReturnsBuyingWinners1993}, this bias tilts stock selection toward sectors that were recent winners and away from recent losers, replicating the logic of an industry momentum strategy. In the LSTM from \textcite{fischerDeepLearningLong2018}, this industry momentum is not captured since all stocks are treated homogeneously because all stocks are trained and classified with the exactly same parameters. 

\begin{figure}
\centering
\includegraphics[width=1\linewidth]{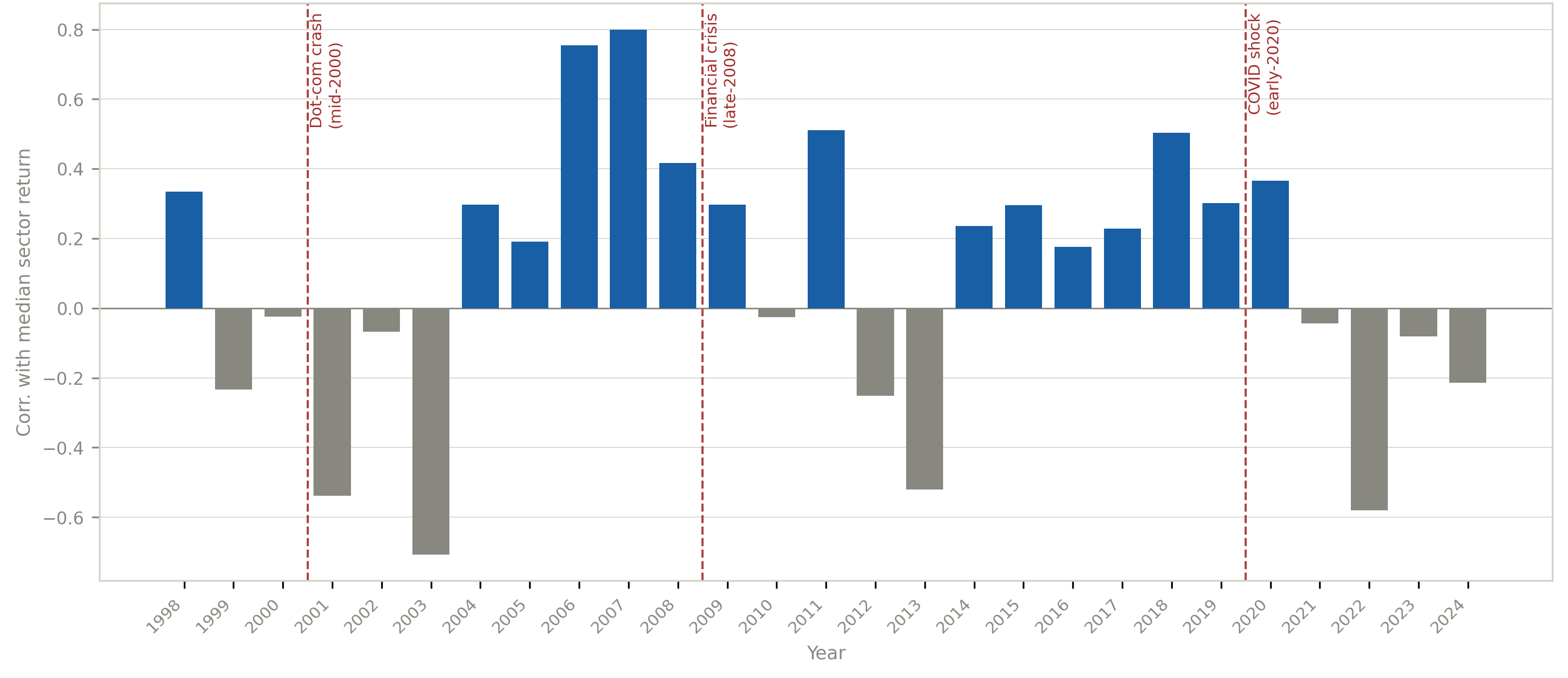}
\caption{\label{fig:momentum_crashes}Correlation of the $\Delta_c$ with the realized median sector return by year.}
\end{figure}

Figure \ref{fig:momentum_crashes} shows the correlation of the $\Delta_c$ with the median sector return. It can be seen that periods of multiple years exist where the correlation is positive, e.g. from 2004 to 2009 and from 2014 to 2020. In these years, the industry momentum captured by $\Delta_c$ helps the Sector LSTM to outperform the base LSTM. Conversely, there are also periods where the correlation is negative for multiple years, most prominently from 1999 to 2003 and from 2021 to 2024. What is striking is that these periods lag major stock crashes. The Dot-com crash from 2000 was at the beginning of the first period of extended negative correlation. The Financial Crisis from 2008 was before the negative years from 2010 up to 2013 and the Covid-19 crash from 2020 was before the negative period from 2021 - 2024. In their paper on momentum crashes, \textcite{danielMomentumCrashes2016} find a similar pattern for periods where the momentum factor performs poorly. In particular, 14 of the 15 worst monthly momentum returns happen when the two-year lagged market return is negative. Among those were months that followed the dot-com crash and the financial crisis respectively.

Additionally, \textcite{danielMomentumCrashes2016} also explicitly mention the period from 2009 to 2013 as one of the largest drawdown periods for momentum strategies (with 2011 being the best year within this period similar to my results). 

\begin{figure}
\centering
\includegraphics[width=1\linewidth]{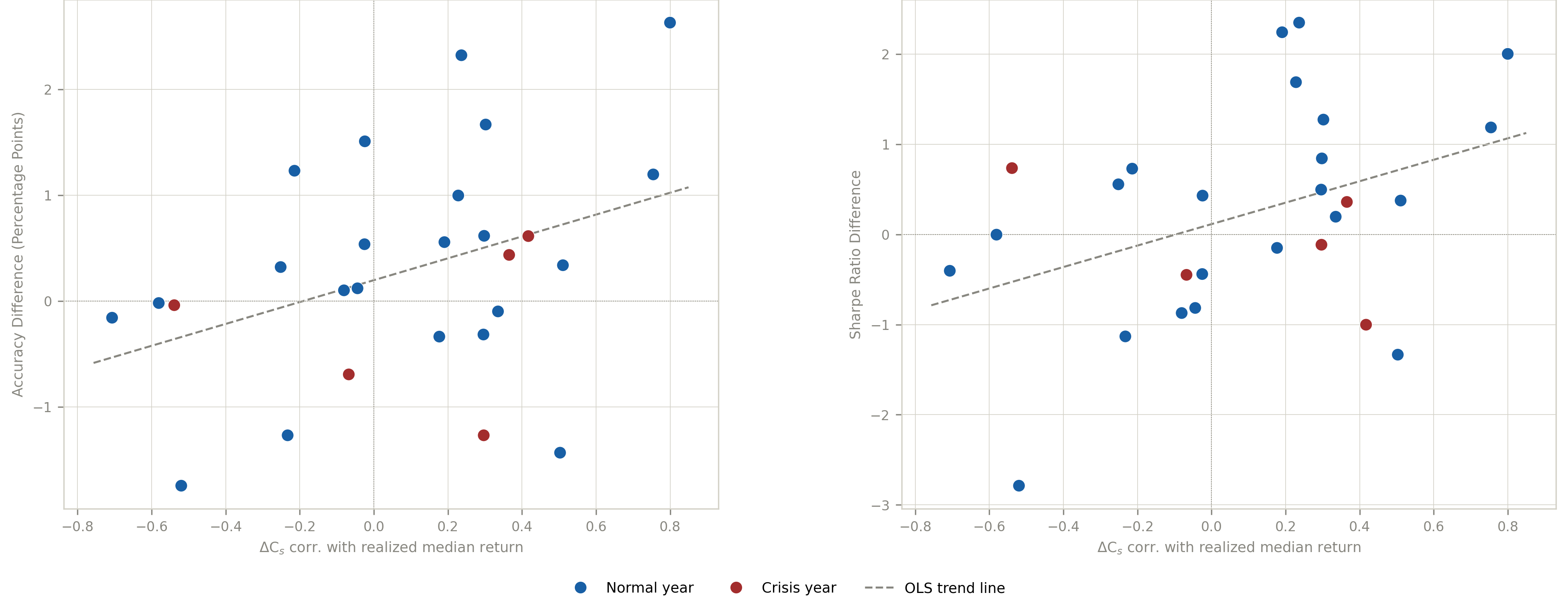}
\caption{\label{fig:embedding_quality}Impact of industry momentum on out-performance of sector LSTM over base LSTM}
\end{figure}

\begin{figure}
\centering
\includegraphics[width=0.9\linewidth]{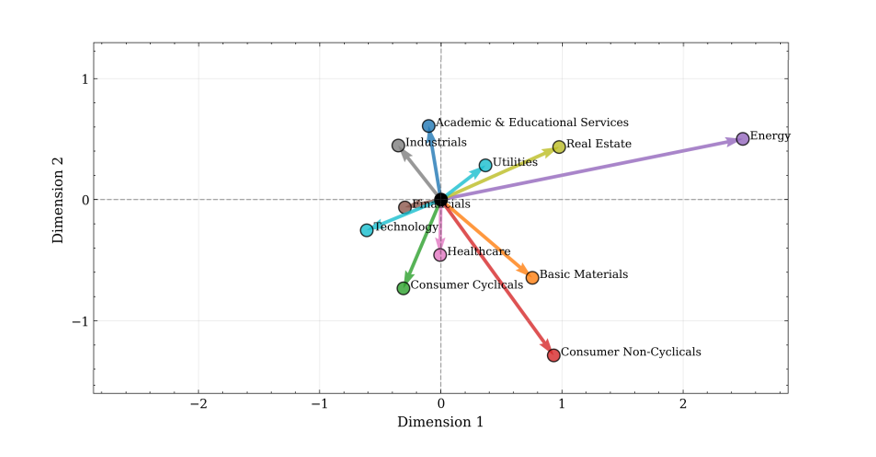}
\caption{\label{fig:sector07weights}Visualization of Sector Embedding Vectors for 2007.}
\end{figure}

The good performance of the sector LSTM from 2004 until 2009 can also be seen in Figure \ref{fig:longtermMetrics}, where the green line (sector LSTM) is above the red line (base LSTM) for the whole period. To assess whether the cross-sectional sector signal improves performance, I test the association between the yearly $\Delta_c$–realized-return correlation and the sector LSTM's out-performance over the base LSTM, measured separately as the difference in accuracy and the difference in annualized Sharpe ratio across the 27 yearly test periods. For each metric I compute the Pearson correlation $r$ between the two series. 
Significance is assessed with a one-sided permutation test, which avoids the bivariate-normality assumption of the parametric correlation test \parencite{goodPermutationTests2000}. Under the null of no association the two series are exchangeable: I draw $B=20,000$ permutations of the performance-difference series, recompute $r$ for each, and estimate the p-value as the proportion of permuted correlations at least as large as the observed value, using the bias-corrected estimator $(\#\{r^* \ge r_{\text{obs}}\} + 1)/(B + 1)$ \parencite{phipsonPermutationPvaluesShould2010}. The one-sided alternative $r>0$ reflects the directional hypothesis that the sector embeddings help precisely when their learned ordering aligns with realized sector returns. Because two hypotheses are tested jointly, I apply the \textcite{holmSimpleSequentiallyRejective1979} step-down correction to control the family-wise error rate.
The Pearson correlations are 0.38 (p-value of 0.025) for the accuracy difference and 0.4 (p-value of 0.019) for the Sharpe Ratio difference, with Holm-adjusted p-values of 0.038. This provides statistical significant evidence that the sector LSTM's out-performance over the base LSTM is associated with the alignment between the learned $\Delta_c$ and realized sector returns. Figure \ref{fig:embedding_quality} visualizes this relationship and additionally plots the OLS fit.

\subsubsection{Case Study: Outperformance in Year 2007}

The 2007 forecasting period provides a positive illustration of the mechanism, as it is a year in which the Sector-LSTM clearly outperforms the base LSTM across all performance metrics (see Figure \ref{fig:longtermMetrics}). The hypothesis that was validated in the previous section is that the sector embedding is the only architectural difference between the two models. Therefore, any systematic out-performance attributable to cross-sectional stock selection must originate from the learned $\Delta_c$ values.

Table \ref{tab:sector_embeddings_07} depicts $\Delta_c$ and actual return measures from the year 2007. To assess the adequacy of the sector contribution, I calculate the correlation of $\Delta_c$ and the various return measures weighted by the number of stocks in each sector:

$$\rho_{xy,w}=\frac{\text{cov}(x,y;w)}{\sqrt{\text{var}(x;w)\text{var}(y;w)}}$$
where the cross-sectional weighted covariance is:
$$\text{cov}(x,y;w)=\frac{\sum_{c=1}^{C}w_{c}(x_{c}-\bar{x}_{w})(y_{c}-\bar{y}_{w})}{\sum_{c=1}^{C}w_{c}}$$
and the weighted mean is:
$$\bar{x}_{w}=\frac{\sum_{c=1}^{C}w_{c}x_{c}}{\sum_{c=1}^{C}w_{c}}$$
with $w_c$ being the number of stocks in sector c, $x_c$ the realized return measure for sector c and $C$ is the number of sectors. The term $\sum_{c=1}^{C}w_{c}$ can be normalized to 1 to simplify the estimation of the weighted first and second moments.

\begin{table}[ht]
\centering
\caption{Comparison of sector contributions and actual return characteristics by sector (all return measures in \%) for 2007}
\label{tab:sector_embeddings_07}
\setlength{\tabcolsep}{4pt}
\footnotesize
\begin{tabular}{l*{7}{r}}
\hline
Sector & \shortstack[r]{Train Median\\Return} & $\Delta_c$ & \shortstack[r]{Mean\\Return} & \shortstack[r]{Median\\Return} & \shortstack[r]{Pct\\Positive} & \shortstack[r]{Ann.\\Return} & \shortstack[r]{Num\\Stocks} \\
\hline
Energy & 0.16 & 0.14 & 0.16 & 0.19 & 54.30 & 36.88 & 31 \\
Real Estate & 0.12 & 0.07 & $-0.01$ & $-0.05$ & 48.40 & $-16.09$ & 16 \\
Utilities & 0.06 & 0.03 & 0.04 & 0.10 & 52.95 & 4.07 & 29 \\
Basic Materials & 0.03 & 0.01 & 0.05 & 0.06 & 51.86 & 3.51 & 26 \\
Industrials & 0.02 & 0.00 & 0.04 & 0.06 & 51.32 & 7.79 & 64 \\
Consumer Non-Cyclicals & 0.01 & $-0.02$ & 0.04 & 0.04 & 50.97 & 4.22 & 48 \\
Financials & 0.02 & $-0.02$ & $-0.08$ & $-0.02$ & 48.55 & $-18.64$ & 79 \\
Healthcare & 0.00 & $-0.02$ & 0.02 & 0.00 & 50.34 & 8.32 & 51 \\
Technology & 0.00 & $-0.04$ & $-0.01$ & 0.00 & 49.46 & $-2.07$ & 82 \\
Consumer Cyclicals & 0.00 & $-0.05$ & $-0.06$ & $-0.04$ & 48.26 & $-19.13$ & 72 \\
\hline
\shortstack[l]{Corr. with $\Delta_c$ (2007)} & 0.97 & 1.00 & 0.74 & 0.80 & 0.75 & 0.67 & \\
\hline
\end{tabular}
\end{table}

The model assigns the largest positive contributions to Energy ($\Delta_c = 0.14$) and Real Estate ($\Delta_c = 0.07$), reflecting their superior relative performance during the training period, with median daily training returns of 0.16\% and 0.12\% respectively. Overall, the correlation between $\Delta_c$ and train median return is 0.97, indicating that the sector embeddings correctly rank the sectors based on the performance in the 3 years of training data. The results are mixed but largely positive in the test year. Energy strongly confirms the sector signal, recording the highest mean daily return across all sectors at 0.16\% (36.88\% annualized). Real Estate has the second-largest positive embedding contribution, but recorded a mean daily return of $-0.01\%$ and an annualized return of $-16.09\%$ in 2007 The divergence can be interpreted as the emerging housing market distress began to weigh on the sector precisely during this period. At the opposite end, Consumer Cyclicals ($\Delta_c = -0.05$) and Technology ($\Delta_c = -0.04$) were assigned the largest negative contributions, consistent with their weak realized returns of $-0.06\%$ and $-0.01\%$ respectively. Notably, Financials with a moderate negative contribution of $\Delta_c= -0.02\%$ recorded the weakest mean daily return of all sectors at $-0.08\%$ ($-18.64\%$ annualized), foreshadowing the sector's broader deterioration ahead of the financial crisis. The weighted cross-sectional correlation between $\Delta_c$ and realized mean daily return across sectors is 74\% for 2007, with correlations of 75\% and 67\% for the percentage of positive return days and annualized return respectively. These figures suggest a moderate to strong positive relationship between learned sector embeddings and return in the test year and thereby confirm the industry momentum for this given period. A visualization of the sectors in the 2-dimensional embedding space can be seen in Figure 13. The values for the sectors correspond to $e_{1,c}$ and $e_{2,c}$ from Figure \ref{fig:finalLayer}.

\begin{figure}
\centering
\includegraphics[width=0.9\linewidth]{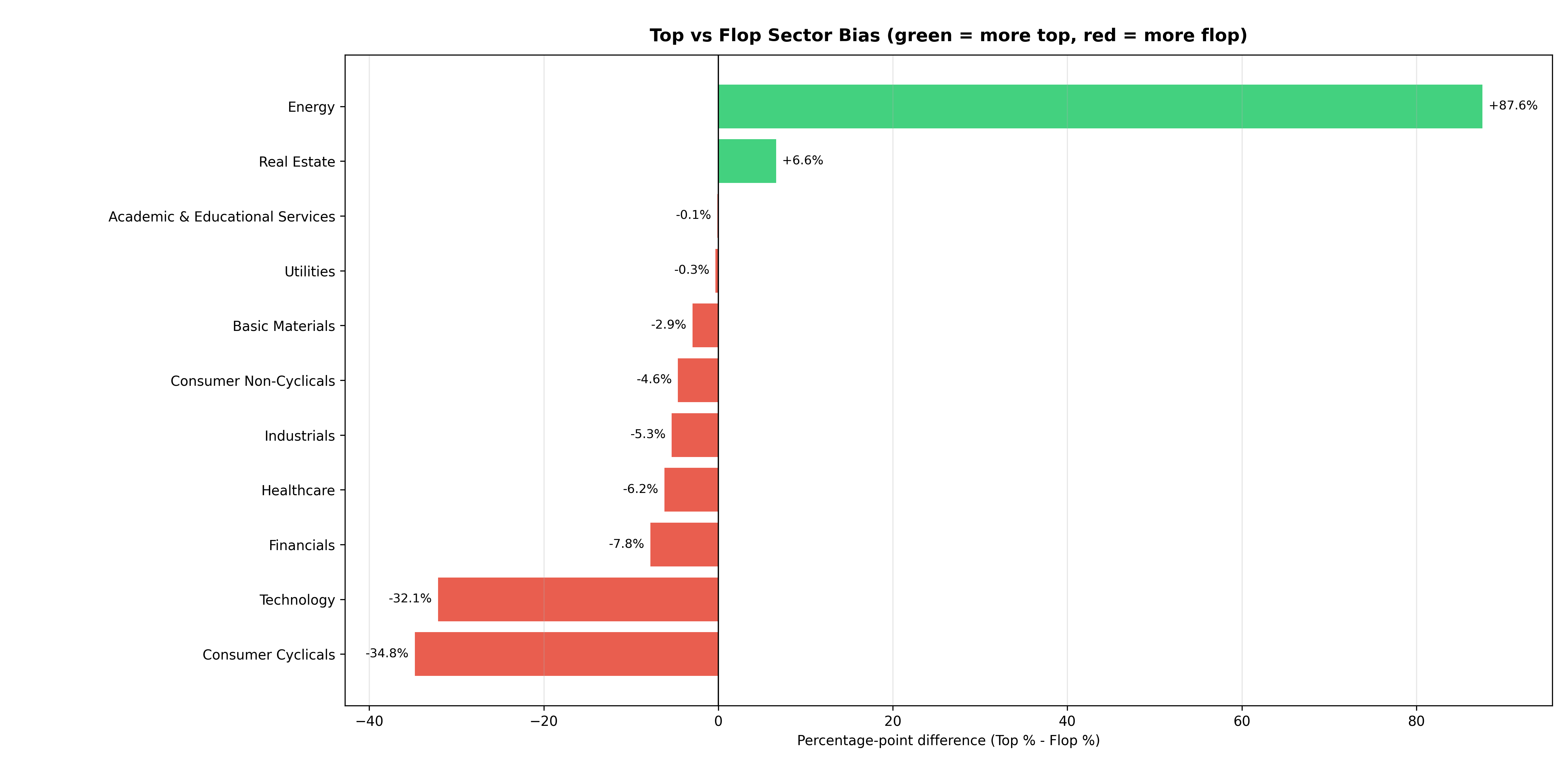}
\caption{\label{fig:stockselection07}Impact of the learned sector embeddings on stock selection into the long short portfolio in 2007.}
\end{figure}

The portfolio-level consequences of these biases are visible in Figure \ref{fig:stockselection07}, which reports the difference in sector representation between the long and short portfolios over the 2007 test period. Energy stocks constitute 87.6 percentage points (pp) more of the long portfolio than the short portfolio, reflecting the dominant positive $\Delta_c$ assigned to that sector. This means that most of the long portfolio consists of Energy stocks showing that these sector biases can have a large effect on stock selection. Conversely, Consumer Cyclicals (-34.8 pp), Technology (-32.1 pp), and Financials (-7.8 pp) are overrepresented in the short portfolio, consistent with their negative embedding values. The sector embeddings thus operate as an active cross-sectional allocation mechanism, systematically steering the long leg of the portfolio toward recent industry winners and away from recent losers.

\subsubsection{Case Study: Momentum Crash Year 2003}

The 2003 forecasting period provides a contrasting illustration of the sector embedding mechanism during a momentum crash episode. As shown in Table \ref{tab:sector_returns_03}, the weighted correlation between the median sector return during training and $\Delta_c$ is 0.86, confirming that the model successfully encoded the cross-sectional return structure of the preceding training window. However, the correlation between $\Delta_c$ and all realized return measures in the test year is strongly negative, ranging from $-0.04$ for the percentage of positive return days to $-0.91$ for annualized total return. For this period, the embedding signal is a misleading indicator for realized returns by industries.

\begin{table}[ht]
\centering
\caption{Comparison of sector contributions and actual return characteristics by sector (all return measures in \%) for 2003}
\setlength{\tabcolsep}{4pt}
\footnotesize
\begin{tabular}{l*{7}{r}}
\hline
Sector & \shortstack[r]{Train Median\\Return} & $\Delta C_{s}$ & \shortstack[r]{Mean\\Return} & \shortstack[r]{Median\\Return} & \shortstack[r]{Pct\\Positive} & \shortstack[r]{Total\\Return} & \shortstack[r]{Num\\Stocks} \\
\hline
Utilities & 0.02 & 0.05 & 0.13 & 0.05 & 51.31 & 15.09 & 33 \\
Healthcare & 0 & 0.03 & 0.12 & 0.09 & 52.09 & 27.04 & 51 \\
Consumer Non-Cyclicals & 0 & 0.03 & 0.07 & 0.04 & 50.79 & 15.81 & 43 \\
Basic Materials & $-0.04$ & 0.02 & 0.15 & 0.10 & 52.41 & 31.50 & 29 \\
Financials & 0 & 0 & 0.12 & 0.07 & 51.83 & 27.72 & 82 \\
Industrials & $-0.04$ & 0 & 0.14 & 0.04 & 50.82 & 31.51 & 70 \\
Energy & 0 & $-0.01$ & 0.13 & 0.07 & 51.13 & 24.75 & 25 \\
Real Estate & 0 & $-0.03$ & 0.14 & 0.11 & 53.40 & 30.05 & 7 \\
Consumer Cyclicals & $-0.02$ & $-0.04$ & 0.14 & 0.07 & 51.26 & 32.50 & 71 \\
Technology & $-0.24$ & $-0.08$ & 0.22 & 0.14 & 51.64 & 53.73 & 87 \\
\hline
\shortstack[l]{Corr. with $\Delta_c$ from 2003} & 0.86 & 1 & $-0.82$ & $-0.71$ & $-0.04$ & $-0.91$ & \\
\hline
\end{tabular}
\label{tab:sector_returns_03}
\end{table}

The economic mechanism underlying this failure is a sharp sector rotation following the dot-com bust. The model assigned its largest positive contributions to Utilities ($\Delta_c = 0.05$) and Healthcare ($\Delta_c = 0.03$), sectors that had exhibited relative resilience during the preceding bear market, and its largest negative contributions to Technology ($\Delta_c = -0.08$) and Consumer Cyclicals ($\Delta_c = -0.04$), the primary casualties of the crash. In 2003, however, the market entered a broad recovery phase in which prior losers rebounded most sharply. Technology recorded the highest total return of any sector at 53.73\%, while Utilities and Healthcare, the model's favored long sectors, returned only 15.09\% and 27.04\% respectively. As a consequence, the sector embeddings systematically directed the long-short portfolio toward the worst-performing sectors and away from the best-performing ones, inverting the intended allocation signal.

\begin{figure}
\centering
\includegraphics[width=0.9\linewidth]{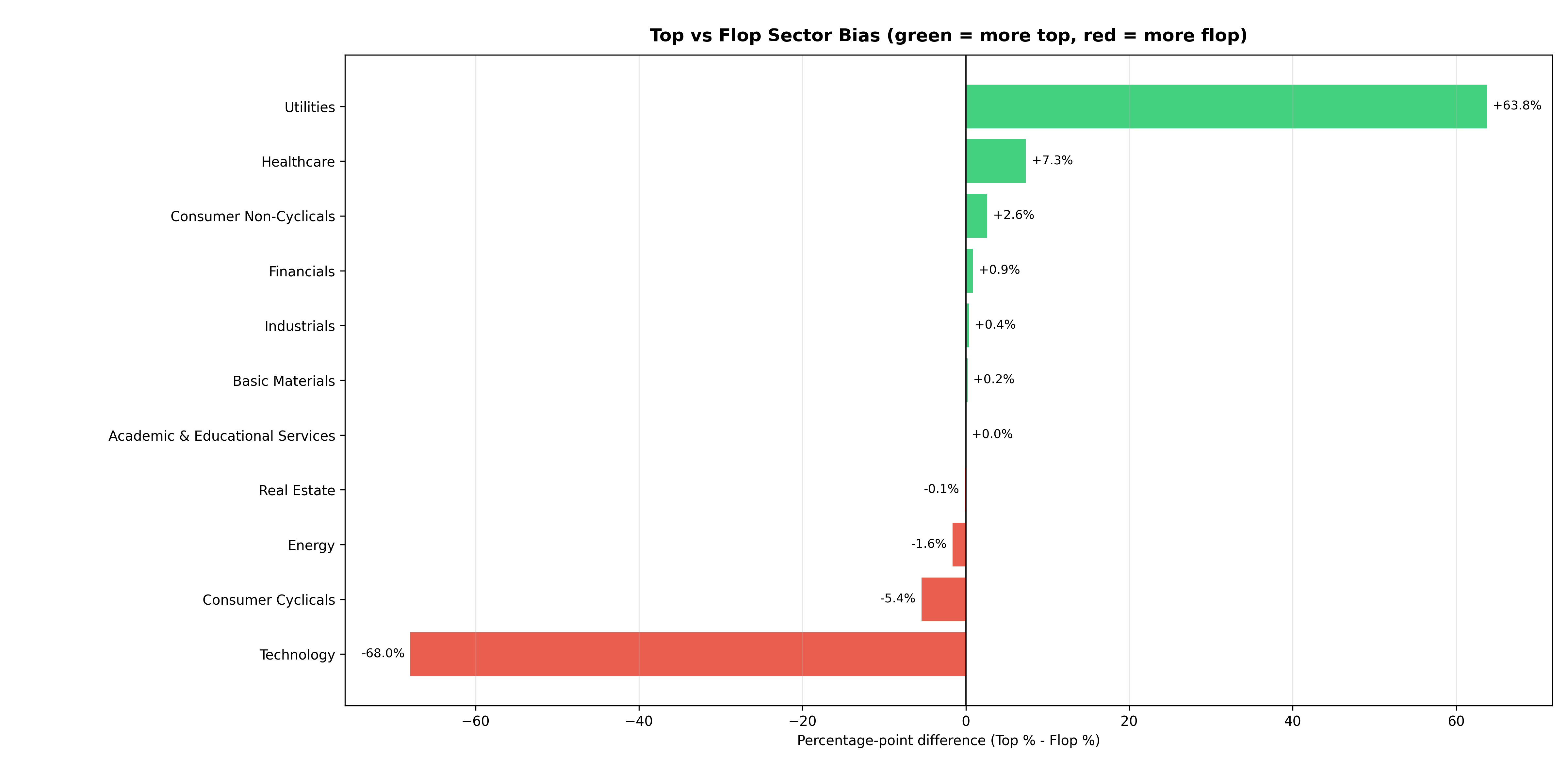}
\caption{\label{fig:stockselection03}Impact of the learned sector embeddings on stock selection into the long short portfolio in 2003.}
\end{figure}

Figure \ref{fig:stockselection03} quantifies the resulting portfolio distortion. Utility stocks constitute 63.8 percentage points more of the long portfolio than the short portfolio, and Healthcare a further +7.3 pp, while Technology stocks are concentrated in the short portfolio with a bias of $-68.0$ pp. This configuration is precisely the one most exposed to a momentum crash: the short leg appreciates substantially while the long leg only has small returns, generating negative overall returns. This mechanism aligns with the findings of Daniel and Moskowitz (2016), who identify the outperformance of prior losers, held in the short portfolio, as the primary driver of momentum crash losses.

\subsubsection{Short-Term Reversal from the Temporal Signal}

Similar to the results in \textcite{fischerDeepLearningLong2018}, the temporal signal from the LSTM shows clear signs of short-term reversal. The phenomenon that stocks with strong returns over the prior week tend to underperform in the subsequent week, and vice versa, was first documented by \textcite{jegadeesh1990evidence} and \textcite{lehmannFadsMartingalesMarket1990}. \textcite{jegadeesh1990evidence} shows that a strategy selecting stocks based on their prior-month returns generates abnormal returns of about 2\% per month over 1965–1989, while \textcite{lehmannFadsMartingalesMarket1990} documents that portfolios of prior-week losers outperform prior-week winners at the weekly horizon. \textcite{loMackinlay1990} attribute a substantial part of these profits to lead-lag cross-autocorrelations across stocks rather than pure own-stock reversal, and subsequent work links the effect to short-term liquidity provision and price pressure that reverses as inventory is absorbed \parencite{nagel2012evaporating}. This literature establishes short-horizon reversal as a robust, if partly liquidity-driven, feature of individual stock returns, in contrast to the medium-horizon momentum that operates at the industry level.

Figure 16 illustrates the average cumulative return of the 10 stocks picked to be in the long and short portfolio each, compared to the average cumulative return of all constituents of the S\&P 500. Roughly five days before the day of choosing the long and short picks, there are remarkable trends in the returns of the top and flop stocks. The candidates with the highest probability to outperform the median S\&P 500 return have seen negative returns in each of the prior days on average. Conversely, the short candidates drastically outperformed the mean of the stock universe in the five days before the decision. Going long recent losers and short recent winners at the daily horizon is exactly the reversal signature documented by \textcite{jegadeesh1990evidence} and \textcite{lehmannFadsMartingalesMarket1990}, recovered here endogenously from the LSTM's temporal component.
\begin{figure}
\centering
\includegraphics[width=0.9\linewidth]{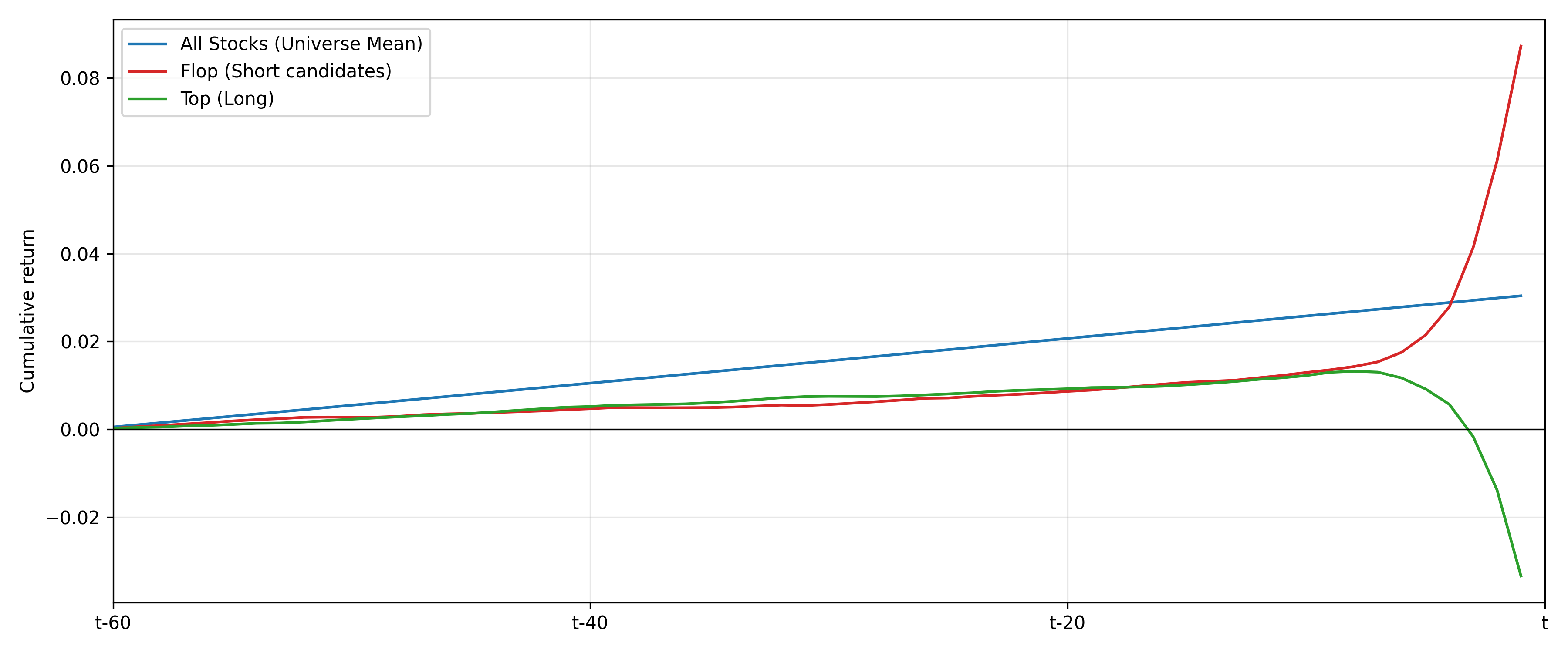}
\caption{\label{fig:reversal}Average cumulative returns of the top and flop 10 stocks picked for the Long-Short Portfolio.}
\end{figure}

\bigskip

To summarize the findings in chapter 4.3, the sector LSTM coincides with the finding by \textcite{jegadeeshMomentum2011} that the momentum of individual stocks is defined by short horizon return reversals while the industry momentum benefits from auto-correlation in portfolio returns.

The strength of the industry momentum signal is not uniform over the full sample. While the weighted correlation between $\Delta_c$ and various realized return measures reaches \>67\% in 2007, the corresponding number averaged across all 27 forecasting periods is only 7.6\%. This can be explained that industry momentum is well-documented to be regime-dependent: it is most profitable during trending markets and substantially weaker or negative during sharp reversals, as demonstrated by the momentum crashes of 2001–2002 and 2009-2013 \parencite{danielMomentumCrashes2016}. In periods characterized by abrupt sector rotation, such as the recoveries following the dot-com bust and the 2008 financial crisis, prior sector winners tend to become underperformers, undermining the persistence assumption that the sector embedding implicitly relies upon.

This interpretation is consistent with the year-by-year performance patterns shown in Figure \ref{fig:longtermMetrics}. The Sector-LSTM's margin of outperformance over the base LSTM is largest in years of relatively stable cross-sectional return dispersion, where sector-level persistence is more likely to hold over a one-year horizon. In crisis years such as 2008 and 2020, both models benefit from elevated volatility in the market regardless of sector, and the incremental value of the embedding signal is reduced. The lower average correlation across all years thus reflects the conditional nature of industry momentum rather than the absence of a systematic effect, and is consistent with the broader momentum literature's finding that the factor premium is time-varying and crash-prone.

\subsection{Robustness}

To assess whether the main results depend on the specific architectural configuration reported in Section 4.1, Table~\ref{tab:robustness_lstm} reports the long-short strategy (with $k=10$) across the four LSTM variants: base, sector, covariate, and a combined sector-plus-covariate LSTM model. Each is estimated at several depths ranging from one to ten five layers and different hyperparameters . All models, together with their fitted parameters, are available on GitHub. Reporting multiple specifications of each model rather than a single preferred specification indicates robustness of the reported results in this section. In the following, I will analyze the most important aspects of the robustness table \ref{tab:robustness_lstm}.

\begin{table}[htbp]
\centering
\caption{Robustness of the long-short strategy ($k=10$) across LSTM architectures. All figures are pooled over the 27 study periods (1995--2024).}
\label{tab:robustness_lstm}
\small
\begin{tabular}{lcccccccc}
\toprule
Model with Layers  & Acc$_{k=10}$ & Acc$_{all}$ & $\bar{r}$ & $t_{NW}$ & SR & $\bar{r}^{tc}$ & SR$^{tc}$ & $\overline{SR}_{y}$ (sd) \\
\midrule
\multicolumn{9}{l}{\textit{LSTM Base}} \\
\quad 1  & 0.519 & 0.505 & 0.082 & 6.26 & 1.10 & 0.023 & 0.22 & 1.17 (1.93) \\
\quad 2  & 0.523 & 0.505 & 0.085 & 6.34 & 1.10 & 0.028 & 0.30 & 1.05 (1.46) \\
\quad 3$^{\dagger}$  & 0.522 & 0.505 & 0.091 & 7.10 & 1.26 & 0.033 & 0.39 & 1.16 (1.43) \\
\quad 5  & 0.518 & 0.504 & 0.081 & 6.09 & 1.08 & 0.024 & 0.24 & 1.04 (1.54) \\
\addlinespace
\multicolumn{9}{l}{\textit{LSTM Sector}} \\
\quad 1 & 0.523 & 0.506 & 0.088 & 6.59 & 1.15 & 0.031 & 0.34 & 1.27 (1.96) \\
\quad 2 & 0.523 & 0.505 & 0.091 & 7.28 & 1.29 & 0.037 & 0.45 & 1.35 (1.77) \\
\quad 3$^{\dagger}$ & 0.525 & 0.506 & 0.100 & 7.81 & 1.41 & 0.053 & 0.69 & 1.39 (1.59) \\
\quad 5 & 0.521 & 0.505 & 0.083 & 7.16 & 1.26 & 0.041 & 0.55 & 1.11 (1.55) \\
\addlinespace
\multicolumn{9}{l}{\textit{LSTM Cov}} \\
\quad 1 & 0.524 & 0.504 & 0.084 & 5.70 & 0.99 & 0.028 & 0.27 & 1.07 (1.59) \\
\quad 2 (without vix) & 0.518 & 0.503 & 0.084 & 5.37 & 0.96 & 0.034 & 0.33 & 0.69 (1.59) \\
\quad 2 & 0.520 & 0.502 & 0.077 & 5.65 & 0.97 & 0.030 & 0.31 & 0.85 (1.53) \\
\quad 3  & 0.519 & 0.503 & 0.071 & 4.89 & 0.83 & 0.020 & 0.16 & 0.78 (1.52) \\
\addlinespace
\multicolumn{9}{l}{\textit{LSTM Cov+Sector}} \\
\quad 1 & 0.523 & 0.505 & 0.085 & 6.57 & 1.15 & 0.029 & 0.32 & 1.15 (1.97) \\
\quad 2 & 0.523 & 0.504 & 0.093 & 6.63 & 1.18 & 0.044 & 0.50 & 1.19 (1.52) \\
\quad 3 & 0.520 & 0.503 & 0.078 & 5.99 & 1.06 & 0.040 & 0.48 & 1.00 (1.41) \\
\quad 5 & 0.514 & 0.503 & 0.065 & 5.13 & 0.86 & 0.036 & 0.42 & 0.78 (1.55) \\
\bottomrule
\end{tabular}
\begin{minipage}{\textwidth}\vspace{2mm}\footnotesize
Notes: Acc$_{k=10}$ is the directional accuracy of the 20 stocks held each day (10 long, 10 short); Acc$_{all}$ is the classification accuracy over all stocks in the cross-section. $\bar{r}$ is the mean daily return of the long-short portfolio in percent, $t_{NW}$ its Newey-West (1 lag) $t$-statistic, and SR the annualized Sharpe ratio computed on excess returns over the risk-free rate. Superscript $tc$ denotes figures after transaction costs of 2 bps per trade. $\overline{SR}_{y}$ is the mean of the 27 period-specific annualized Sharpe ratios, with the standard deviation across periods in parentheses. $^{\dagger}$ marks the specification reported in the main results.
\end{minipage}
\end{table}

First, directional accuracy is stable across all architectures and depths: the accuracy of the traded portfolio ($Acc_{k=10}$) lies within the narrow band of 0.514 to 0.525, and the cross-sectional classification accuracy ($Acc_{all}$) between 0.502 and 0.506. The predictive signal is therefore not an artifact of a particular specification of the LSTM models. Given the low signal-to-noise ratio of daily returns, the stability across sixteen specifications is reassuring.

Second, and more importantly for the paper's central claim, the ranking of architectures is preserved across depths. At every layer count the sector LSTM matches or exceeds the base LSTM on mean return and Sharpe ratio, while the covariate LSTM is dominated by both. The combined sector-plus-covariate model does not improve on the sector model alone, reinforcing the earlier conclusion that the macro-financial covariates contribute no exploitable directional signal and, if anything, dilute the sector effect. That the sector embedding delivers a consistent edge regardless of the depth of the neural network is direct evidence that the gain originates in the cross-sectional embedding rather than in an incidental interaction with a specific configuration.

Third, while accuracy and the architecture ranking are robust, the magnitude of the risk-adjusted performance does vary with depth. Within each architecture the return and Sharpe metrics are non-monotonic in layer count, typically peaking at intermediate depths and deteriorating at five layers. This is expected rather than anomalous: deeper LSTMs applied to 60-step sequences are prone to optimization difficulties and vanishing gradients (Hu et al., 2019), so the degradation at high depth reflects known training pathologies, not a failure of the underlying signal. The three-layer specification marked with $†$, used in the main results, was selected on validation loss. The qualitative conclusions of the paper hold across the shallow-to-moderate depths (one to three layers) that a practitioner would plausibly deploy.

Fourth, one can see that the Sharpe Ratio before trading costs is generally higher for the base LSTM compared to the covariate LSTM. However, after trading costs, the covariate LSTM has a better or just slightly below Sharpe Ratio compared to the base LSTM. This can be attributed to the reduced trading of the strategy based on the covariate LSTM. If the same stocks are in the long leg in multiple consecutive days, then the trading strategy keeps them in the long leg without closing the position between those days. Hence, lower turnover means less trading costs which benefits the metrics after trading costs. It can be explained by the fact, that the covariates signal for multiple days that certain stocks should stay in the long or short leg of the portfolio and hence resulting in fewer transactions.

Taken together, the robustness check establishes that the paper's conclusions, a stable directional signal, a consistent advantage for sector embeddings, and no benefit from macro-financial covariates, are robust to the choice of network depth and hyperparameters. What varies across specifications is the precise magnitude of returns and Sharpe ratios, consistent with the broader finding that the economic value of the strategy is modest and regime-dependent.

\section{Conclusion}
This paper extended the return-only LSTM of \textcite{fischerDeepLearningLong2018} along three architectural dimensions: 1) macro-financial covariates, 2) extensive regularization and 3) learnable TRBC sector embeddings. Subsequently, the resulting models were evaluated on a survivorship-bias-free S\&P 500 universe across 27 rolling study periods from 1995 to 2024.

Three findings emerge. First, the sector embedding in combination with the regularization are the extensions that delivers consistent gains: the sector LSTM produces the highest average returns and accuracy among the tested models, a significantly higher Sharpe ratio than both the base and covariate LSTMs and superior predictive accuracy relative to the Random Forest. Although including the static cross-sectional intercept only adds 26 parameters if a two-dimensional sector embedding is used, the sector LSTM consistently outperforms the base LSTM. The macro-financial covariates, by contrast, did not improve directional prediction and degraded risk-adjusted performance, so their broader economic signal offered no exploitable content for this task.

Second, extending the evaluation window through 2024 confirms and lengthens the deterioration documented by \textcite{fischerDeepLearningLong2018}. Strong returns and high Sharpe ratios are concentrated before 2008 with the exception of 2020, no subsequent year approaches the profitability of the late 1990s. This is consistent with the broader literature on diminishing returns to quantitative arbitrage and suggests the post-2010 decline reflects a durable shift rather than a transient phenomenon.

Third, the architecture permits a decomposition of the predictive signal into two economically interpretable components that emerge endogenously from gradient descent. The sector embedding functions as an implicit industry momentum factor: the sector contribution $\Delta_c$ aligns with subsequent realized sector returns, and its year-by-year alignment is significantly associated with the sector LSTM's outperformance over the base LSTM. This alignment is regime-dependent. While it performs strongly in stable years such as 2007, it can have negative impact on returns if there is a sharp sector rotation such as in 2003, mirroring the momentum-crash pattern of Daniel and Moskowitz (2016). The temporal component, in turn, exhibits the short-horizon reversal of Lehmann (1990) which was already documented by \textcite{fischerDeepLearningLong2018}. That two well-documented asset-pricing regularities surface from a binary classification objective offers independent evidence that they are genuine features of the return-generating process.

Taken together, the results suggest that incorporating cross-sectional heterogeneity through learnable embeddings is a parsimonious and interpretable way to improve LSTM-based return prediction, while also cautioning that the incremental signal is modest, conditional on market regime, and has weakened in the last decade.

\newpage

\printbibliography

\newpage

\section{Appendix}

\begin{table}[htbp]
\centering
\caption{Risk and return characteristics}
\label{tab:return-characteristics}
\begin{tabular}{lrrrr}
\toprule
 & Market & LSTM Base & LSTM Sector & Random Forest \\
\midrule
Mean return & 0.049 & 0.033 & 0.053 & 0.041 \\
Mean return (long) & 0.000 & 0.006 & 0.011 & 0.013 \\
Mean return (short) & 0.000 & 0.028 & 0.041 & 0.029 \\
Std.\ dev. & 1.196 & 1.043 & 1.034 & 0.996 \\
Min & -12.000 & -12.704 & -8.248 & -16.723 \\
Max & 11.360 & 12.297 & 10.814 & 10.847 \\
Median & 0.080 & 0.014 & 0.039 & 0.014 \\
Q1 & -0.460 & -0.424 & -0.423 & -0.386 \\
Q3 & 0.620 & 0.449 & 0.497 & 0.457 \\
Share positive & 0.542 & 0.508 & 0.522 & 0.508 \\
Skewness & -0.256 & 0.749 & 0.288 & -0.051 \\
Kurtosis & 9.110 & 15.586 & 9.691 & 24.451 \\
$t$-statistic & 3.657 & 2.615 & 4.106 & 3.406 \\
VaR (1\%) & -3.360 & -2.954 & -2.808 & -2.839 \\
VaR (5\%) & -1.840 & -1.401 & -1.416 & -1.292 \\
CVaR (1\%) & -4.749 & -3.977 & -4.092 & -3.986 \\
CVaR (5\%) & -2.843 & -2.318 & -2.317 & -2.229 \\
Max drawdown & -0.546 & -0.830 & -0.556 & -0.762 \\
Ann.\ return & 0.122 & 0.084 & 0.133 & 0.104 \\
Ann.\ volatility & 18.984 & 16.557 & 16.414 & 15.816 \\
Excess ann.\ return & 0.099 & 0.064 & 0.112 & 0.084 \\
Sharpe ratio & 0.523 & 0.387 & 0.685 & 0.530 \\
\bottomrule
\end{tabular}
\end{table}

\end{document}